\documentclass[twocolumn]{article}
\usepackage{graphicx}
\usepackage{authblk}
\usepackage{hyperref}
\usepackage{footmisc}
\usepackage{xcolor}
\usepackage{amsmath}
\usepackage{geometry}
\usepackage{enumitem}
\usepackage{caption}
\usepackage[doi=true,isbn=false,url=false,eprint=false,style=numeric]{biblatex}

\hypersetup{
    colorlinks=true,
    linkcolor=blue,
    filecolor=black,      
    urlcolor=blue,
    citecolor=black
}

\newcommand{\cotwo}{CO\textsubscript{2}\,}
\newcommand{\dry}{_{\textrm{dry}}}
\newcommand{\wet}{_{\textrm{wet}}}
\newcommand{\soil}{_{\textrm{soil}}}

\title{Mass-Balance MRV for Carbon Dioxide Removal by Enhanced Rock Weathering\\{\Large Methods, Simulation, and Inference}}

\author[1]{Mark Baum\thanks{\href{mailto:mark@lithoscarbon.com}{mark@lithoscarbon.com}}}
\author[1]{Henry Liu}
\author[1]{Lily Schacht}
\author[1]{Jake Schneider}
\author[1]{Mary Yap}

\affil[1]{Lithos Carbon (\href{https://www.lithoscarbon.com/}{\texttt{lithoscarbon.com}})}

\date{\today}

\begin{document}

\maketitle

\begin{abstract}
Carbon dioxide will likely need to be removed from the atmosphere to avoid significant future warming and climate change. Technologies are being developed to remove large quantities of carbon from the atmosphere. Enhanced rock weathering (ERW), where fine-grained silicate minerals are spread on soil, is a promising carbon removal method that can also support crop yields and maintain overall soil health. Quantifying the amount of carbon removed by ERW is crucial for understanding the potential of ERW globally and for building trust in commercial operations. However, reliable and scalable quantification in complex media like soil is challenging and there is not yet a consensus on the best method of doing so. Here we discuss mass-balance methods, where stocks of base cations in soil are monitored over time to infer the amount of inorganic carbon brought into solution by weathering reactions. First, we review the fundamental concepts of mass-balance methods and explain different ways of approaching the mass-balance problem. Then we discuss experimental planning and data collection, suggesting some best practices. Next, we present a software package designed to facilitate a range of tasks in ERW like uncertainty analysis, planning field trials, and validating statistical methods. Finally, we briefly review ways of estimating carbon removal using mass balance before discussing some advantages of Bayesian inference in this context and presenting an example Bayesian model. The model is fit to simulated data and recovers the correct answer with a clear representation of uncertainty.
\end{abstract}

\tableofcontents

\newpage

\section{Introduction \& Background}
\label{sec:intro}

Large-scale removal of carbon dioxide from the atmosphere is probably required to avoid significant future warming and climate change \cite{masson2021ipcc}, but removing large masses of \cotwo from the atmosphere is challenging. Carbon dioxide removal (CDR) on the necessary scale---billions of tons per year---is not currently practical. However, an array of technologies are being developed to accomplish large-scale CDR. One promising method is enhanced rock weathering (ERW), which accelerates the natural silicate rock weathering process that transfers carbon from the atmosphere to the ocean over geologic timescales \cite{hartmann_enhanced_2013, beerling_farming_2018}.

In ERW, finely pulverized silicate materials like basalt and olivine, usually called ``feedstocks," are spread over large areas of soil like cropland and pasture. Feedstock materials dissolve in the presence of naturally occurring carbonic acid in soil, bringing carbon into solution as bicarbonate (HCO$3^{-}$). The disintegration of these minerals is called ``weathering" and the weathering process is thought to regulate global atmospheric \cotwo concentration over very long (geologic) periods of time \cite{walker_negative_1981}. Spreading fine-grained and highly weatherable material over large areas accelerates the natural weathering rate. Plant respiration also increases \cotwo partial pressure in the soil pore space, accelerating weathering reactions.

Bicarbonate ions produced by weathering reactions are generally stable in solution and are transported in groundwater to continental rivers and eventually to the ocean, where they are also stable for long periods of time (generally $>$10 kyr). On geologic time scales, carbon in solution is cemented on the sea floor in carbonate minerals. Alternatively, bicarbonate ions produced in soil may form terrestrial carbonates or become adsorbed on clay minerals or organic material. We refer to \textcite{hartmann_enhanced_2013} and references therein for a more detailed discussion of relevant carbonate chemistry and speciation. 

In addition to removing carbon, ERW can support soil health and crop yields. The dissolution of silicate materials neutralizes soil acidity and increases soil pH, which is often essential to avoid reduced crop yields on intensively farmed agricultural land \cite{weil_ray_r_brady_nature_2016}. In this capacity, silicate feedstocks replace more commonly used agricultural lime. Silicate minerals may also contain appreciable amounts of potassium, an important nutrient for plant growth, and a range of plant micronutrients. The combined effects of carbon removal, soil pH regulation, and nutrient supply constitute a strong argument for ERW.

Quantifying the amount of atmospheric carbon removed by ERW reliably, cheaply, and accurately is critical for our understanding of its carbon removal potential and commercial viability. In the commercial setting, the whole process of CDR quantification is usually called measurement, reporting, and verification (MRV). Because ERW takes place in diverse natural soils, occurs over large areas, and involves complex chemical reaction networks, MRV is hard. It requires well-designed data collection plans, careful operations in the field, laboratory analysis with consistently high standards, competent data management, and rigorous statistics. Given the scope of the problem and the variety of possible ways to approach MRV, there isn't yet a clear consensus on how best to quantify CDR in ERW.

Importantly, because the products of silicate weathering reactions persist for a long time and may participate in further reactions far away from the original location, MRV is primarily concerned with the \emph{initial} CDR: that which takes place upon feedstock dissolution in the top layer of amended soil. This document focuses entirely on \emph{initial} CDR and its quantification. Downstream loss of removed carbon and its uncertainty is accounted for with models, which suggest a relatively small impact \cite{zhang_river_2022}, but this is also an active area of research.

Several different approaches to MRV in ERW have been explored. Here, we address one promising group of approaches: mass-balance methods.
\begin{itemize}
    \item In Section \ref{sec:mass-balance}, we review fundamental concepts of mass-balance MRV and discuss the strengths and weaknesses of different variations on the core mass-balance approach. We focus mainly on tracer methods because some assumptions are not always obvious and some important caveats have been overlooked.
    \item In Section \ref{sec:planningexperiments}, we review some best practices and suggest ways of avoiding bias and error as different MRV ideas are tested.
    \item In Section \ref{sec:simulation}, we present a software package \cite{baum_2024_11621611} designed specifically to facilitate tasks like sample planning, uncertainty analysis, power analysis, and statistical validation in mass-balance MRV. To demonstrate the package's core functionality, we perform sensitivity analyses on soil-feedstock mixing scenarios.
    \item In Section \ref{sec:estimationinference}, we briefly review some approaches to CDR estimation and review some advantages of Bayesian modeling for mass-balance MRV. Then we describe and fit a demonstration Bayesian model using simulated data from our new package. We show that the model recovers the correct answer for CDR, in addition to other information.
    \item Finally, in Section \ref{sec:conclusion}, we review some of the main points in previous sections and conclude.
\end{itemize}

This is a practical document focusing on how to calculate initial CDR using mass balance methods. It is partly a review of existing research, partly an explanation of some overlooked components of mass-balance methods, partly a resource for building intuition, and partly a presentation of new computational tools for MRV in ERW. It is addressed primarily to researchers and commercial practitioners in ERW, but also organizations seeking to develop standards and protocols for carbon crediting. This document is not, however, an assertion that any individual protocol for MRV is optimal. It also does not address every component of ERW science completely. In particular, the details of silicate weathering reactions, carbonate chemistry, and life cycle analysis are not addressed here. 

\section{Mass-Balance MRV}
\label{sec:mass-balance}

Initial carbon dioxide removal by enhanced rock weathering is driven by the reaction of carbonic acid with silicate minerals to produce bicarbonate anions and base cations. To quantify initial CDR in ERW, we have to observe a causal relationship between feedstock application and one or more of the following:
\begin{itemize}[itemsep=1pt,parsep=1pt]
    \item increased net \cotwo flux from the atmosphere to the soil
    \item elevated bicarbonate concentration in soil porewater
    \item faster dissolution and leaching of base cations (mass-balance)
\end{itemize}
Each of these measurement targets has strengths and weaknesses.

Measuring net \cotwo is a very direct way to quantify CDR because removing \cotwo from the atmosphere is the ultimate goal. It requires no further assumptions about weathering reactions. However, gas phase measurements require continuous operation of well-controlled sensing apparatuses with very high temporal resolution \cite{paessler_monitoring_2023}. This method may be useful in laboratory and mesocosm experiments but is very unlikely to be widely deployed.

Bicarbonate measurement is also relatively direct but is similarly difficult to perform at scale. Aqueous phase measurements require frequent collection of leachate from amended soils and leachate collection is a laborious task, requiring the installation of either suction lysimeters coupled with estimates of precipitation and evapotranspiration or expensive gravity lysimeters that disturb the soil \cite{holzer_direct_2023}.

Measuring feedstock dissolution and leaching only requires occasional collection of samples from amended soil, a practice that is familiar to most farmers, who test soil properties like acidity and nutrient availability. These samples can be digested and analyzed for elemental composition in a centralized, off-site laboratory. However, these measurements ultimately require assumptions about which chemical reactions occur and how much carbon is brought into solution by feedstock dissolution.

There are several possible ways to measure feedstock dissolution and leaching, collectively called ``mass-balance" methods \cite{clarkson2023review}. The goal of any mass-balance method is to estimate the mass of cations dissolved from applied feedstock as a proxy for bicarbonate production and \cotwo removal. Doing so requires, at minimum, information about:
\begin{enumerate}
    \item The mass of feedstock cations in the soil after adding feedstock to soil
    \item The mass of feedstock cations in the soil after a period of weathering
\end{enumerate}
The difference between these masses is the loss of cations from feedstock. Importantly, cation mass is only ``lost" if it both dissolves from feedstock and leaches into deeper soil, where it is no longer within the sampling depth.

Cation mass losses are generally converted to initial CDR by assuming two moles of bicarbonate are brought into solution for every mole of dissolved divalent cations (usually calcium and magnesium) and one mole for every mole of dissolved monovalent cations (usually sodium and potassium). Then, the mass of removed \cotwo is calculated using a 1:1 ratio between moles of bicarbonate produced and \cotwo removed. In particular, for known masses of cations dissolved from silicate feedstock, the mass conversion formula is:
\begin{equation}
    \textrm{\cotwo} = 3.62\,\textrm{Mg} + 2.2\,\textrm{Ca} + 1.91\,\textrm{Na} + 1.12\,\textrm{K} \, .
    \label{eq:cation2co2}
\end{equation}
For example, one kilogram of magnesium dissolved from feedstock implies 3.62 kilograms of initial CDR. This is sometimes expressed differently, with oxides, more elements, and various loss factors, and called the Steinour Formula \cite{renforth_negative_2019}, but it's fundamentally a molar mass conversion. The conversion itself represents the maximum amount of carbon initially removed by the dissolution of feedstock in soil. As mentioned above, subsequent evasion and its uncertainty must also be considered, along with emissions caused by operations, to estimate net CDR.

Measuring the difference in feedstock cation masses over time can be done in many ways, but there are two categories of methods: tracer methods and cation stock methods. Below, we explain the methods in each category and discuss their strengths and weaknesses.

\subsection{Tracer Methods}
\label{subsec:tracers}

Tracer methods take advantage of the relative stability of different silicate minerals and their chemical compositions. For example, ``high field strength elements" like zirconium or titanium tend to be concentrated in minerals that resist chemical weathering (zircon, for example). A considerable amount of information about long-term soil development can be inferred by assuming these minerals are not broken down over time and the masses of their elemental constituents are conserved in a given section of soil. For example, we assume zirconium mass is conserved over time, acting as a passive ``tracer" element during pedogenesis (soil development) and weathering. Elements like calcium and magnesium, which are generally much more soluble, are ``mobile" elements compared to tracers.

Although inspired by research on natural weathering and pedogenesis, tracer methods for ERW have been approached slightly differently. In landmark pedogenesis studies, the composition of a parent rock is approximately known, and tracer elements are used to infer changes in the volume of soil layers as the parent rock transitions into a soil \cite{chadwick_black_1990, brimhall_quantitative_1991}. For ERW, in contrast, tracers have been used to infer the mass of feedstock originally added to soil samples. In effect, they have been used to reconstruct the composition of the parent material itself---soil composition immediately after applying feedstock \cite{reershemius_initial_2023, reershemius_new_2023, beerling_enhanced_2024, kantola_improved_2023}---without measuring it directly after feedstock spreading.

\begin{figure}
    \centering
    \includegraphics[width=\linewidth]{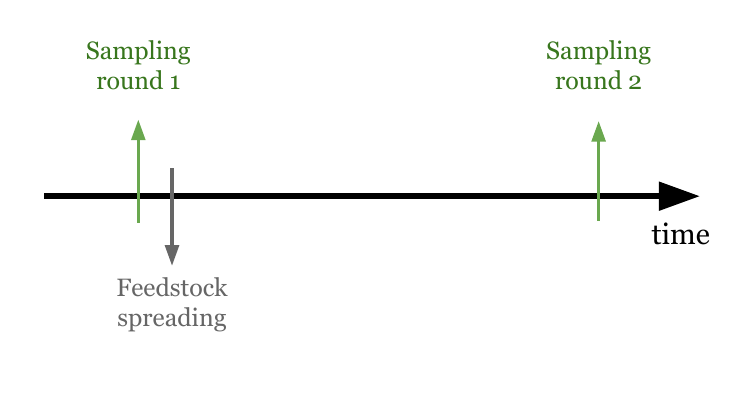}
    \caption{A simple timeline showing sampling rounds before feedstock application and after some period of weathering.}
    \label{fig:prespread_sampling}
\end{figure}

To further explain, tracer methods in ERW have mainly relied on two groups of soil samples in addition to feedstock samples. The first group of soil samples is taken before feedstock is applied and the second is taken some time after weathering has occurred. For example, the first group might be collected one week before spreading and the second group one year after spreading (Figure \ref{fig:prespread_sampling}). All samples are analyzed for chemical composition, recording the mass concentrations of mobile cations and tracer elements\footnote{Typically, samples are prepared for analysis by drying, pulverization, and digestion. Drying and pulverizing are important for homogenization of the sample, but also to optimize particle size, typically smaller than 75 um in diameter, for complete digestion. The two primary methods of digestion are lithium borate fusion or total-acid (sometimes called four-acid) digestion. Samples are then analyzed by X-ray fluorescence (XRF), inductively coupled plasma optical emission spectroscopy (ICP-OES), inductively coupled plasma mass spectrometry (ICP-MS), or a combination of these methods. Feedstock chemical composition is also analyzed in the same way.}.

To summarize, the three groups of measurements are:
\begin{enumerate}[itemsep=1pt,parsep=1pt]
    \item feedstock concentrations ($f$)
    \item soil concentrations before spreading ($s$)
    \item soil concentrations after weathering ($w$)
\end{enumerate}

We assume that none of the tracer element's mass is lost from the soil and that the concentration of a tracer element immediately after applying feedstock is the same as when it's measured a year later. A simple mixing equation describes the concentration of an element in the feedstock-soil mixture,
\begin{equation}
    w_t = \alpha f_t + (1 - \alpha) s_t \,
    \label{eq:mixingi}
\end{equation}
where $\alpha$ is the mass of feedstock mixed into the soil, in units of mass per mass, or the ``mixing fraction." The subscript $t$ indicates the concentrations refer to an immobile tracer element.

Note that the mixing equation above is just a mass-weighted average. It's derived by starting from the masses of each material. If we mix feedstock mass $F$ and soil mass $S$ with elemental concentrations $f_t$ and $s_t$, we have a mixed mass $W$ with some concentration $w_t$,
\begin{equation}
    W w_t = F f_t + S s_t \, .
\end{equation}
Rearranged,
\begin{equation}
    w_t = \frac{F}{W}f_t + \frac{S}{W} s_t \, .
\end{equation}
The fraction $F/W$ is the amount of feedstock mass per unit total mass in the mixture. By definition, this is the mixing fraction $\alpha$, and
\begin{equation}
    w_t = \alpha f_t + \frac{S}{W} s_t \, .
\end{equation}
\textbf{If no mass is lost}, then $W = S + F$. The total mass is just the sum of the soil and feedstock masses. Or, equivalently, $S = W - F$.
\begin{align}
    w_t &= \alpha f_t + \frac{W - F}{W} s_t \\[1ex]
    &= \alpha f_t + \left(1 - \frac{F}{W} \right) s_t \\[1ex]
    &= \alpha f_t + (1 - \alpha_t) s_t
\end{align}
This is where the simple mixing equation comes from. Rearranging, the mixing fraction is
\begin{equation}
    \alpha_t = \frac{w_t - s_t}{f_t - s_t} \, .
    \label{eq:alphai}
\end{equation}

The same mixing algebra applies to any mobile element $m$,
\begin{equation}
    \alpha_m = \frac{w_m - s_m}{f_m - s_m} \, .
\end{equation}
However, for a mobile element, the mixing fraction $\alpha_m$ typically records the soil's state \textit{after} some period of weathering. If all of the feedstock mobile element has been lost, $\alpha_m \approx 0$ because $w_m \approx s_m$. If none of the feedstock mobile element has weathered away, then $\alpha_m = \alpha_t$. The ratio of mixing fractions, $\alpha_m / \alpha_t$, indicates how much of the feedstock mobile element $m$ is present in the soil as a fraction of the amount that was originally added to the soil. This is how tracers can be used to estimate cation losses without measuring their concentrations immediately after spreading. The tracer mixing fraction $\alpha_t$ records the initial amount of feedstock added to the soil and the mobile element mixing fraction $\alpha_m$ records cation loss by comparison with $\alpha_t$.

\subsubsection{Dissolution Fraction}
\label{subsubsec:disfrac}

The relationship between mobile and immobile mixing fractions leads to a dissolution fraction $d$,
\begin{equation}
    d = 1 - \frac{\alpha_m}{\alpha_t} = 1 - \frac{f_t - s_t}{w_t - s_t}\frac{w_m - s_m}{f_m - s_m} \, ,
    \label{eq:dissolutionfrac}
\end{equation}
which is the fraction of the feedstock mobile element $m$ lost from the sampled soil in units of mass per mass. The expression in Equation \ref{eq:dissolutionfrac} can, theoretically, be used to estimate CDR from measurements of each variable. The dissolution fraction for each feedstock cation is multiplied by the mass of the element applied in feedstock, yielding the total mass of the cation lost to weathering, which is then used in Equation \ref{eq:cation2co2} to compute CDR. Note that there is no need to formulate a system of linear equations to compute the dissolution fraction and (as discussed below) statements of mass conservation in this context \cite{reershemius_initial_2023} are misleading and must be qualified.

\subsubsection{Tracer Differencing}
\label{subsubsec:disdif}

Instead of computing a dissolution \emph{fraction}, tracers can also be used to infer the \emph{difference} in mobile element concentrations due to weathering. If feedstock is applied to soil with mixing fraction $\alpha$, then immediately after spreading, the combined concentration of mobile element is
\begin{equation}
    \alpha f_m + (1 - \alpha) s_m \, .
\end{equation}
If we subtract the observed post-weathering concentration,
\begin{equation}
    \alpha f_m + (1 - \alpha) s_m - w_m \, ,
\end{equation}
then we have an expression for how much a mobile element's concentration decreased because of dissolution and leaching. Then, to compute CDR, further assumptions or measurements of density and depth are required because the expression above is a change in cation concentration, not mass.

\subsubsection{Comments \& Caveats}
\label{subsubsec:comcav}

The primary practical strength of tracer methods is that they automatically account for the amount of feedstock added to the observed soil without observing soil concentrations immediately after spreading, even after multiple rounds of spreading. From that estimate of total feedstock application, we can further estimate the expected concentration of mobile elements if no feedstock weathering occurred. Then, observed mobile element concentrations are compared to that estimate. For the dissolution fraction, mobile element loss is \emph{normalized} by its enrichment from feedstock addition. Alternatively, the concentration gap is directly converted to a total mass loss using additional information (``dissolution difference").

There are, however, some important caveats to be aware of when using tracer methods for MRV. Some of them have not been fully acknowledged in prior work.

First, the denominator of the first fraction in Equation \ref{eq:dissolutionfrac}, $w_t - s_t$, can cause severe estimation problems. Natural variability in soil concentrations and imperfect laboratory measurements can cause $w_t$ and $s_t$ to be poorly separated. The difference $w_t - s_t$ can, just by chance, be very close to zero, making the first fraction very large and blowing up the dissolution fraction. It's also possible for the dissolution fraction to be negative or greater than one, outside the expected physical interval. The probability of these unexpected/outlier dissolution fractions increases as $w_t$ and $s_t$ become noisier and if the feedstock is poorly enriched in the tracer element.

However, an important qualification to remember when confronting real data is that dissolution fractions $\notin [0,1]$ \emph{do not} necessarily represent unphysical results. Although they \emph{could} be caused by unknown sources of error, it's more likely that they are caused by natural variability and noise, which will always be present in each term of Equation \ref{eq:dissolutionfrac}. With enough variability, too few samples, and/or unwise analysis choices, values of the dissolution fraction that can't be physically justified are likely to appear. However, noisy results never justify discarding data or clamping dissolution fractions for sample pairs between zero and one \cite{reershemius_initial_2023}, which introduces bias. Dissolution fractions outside [0,1] probably indicate some combination of noisy data, a badly behaved estimand, and insufficient experimental planning, not unphysical behavior.

\begin{figure}
    \centering
    \includegraphics[width=\linewidth]{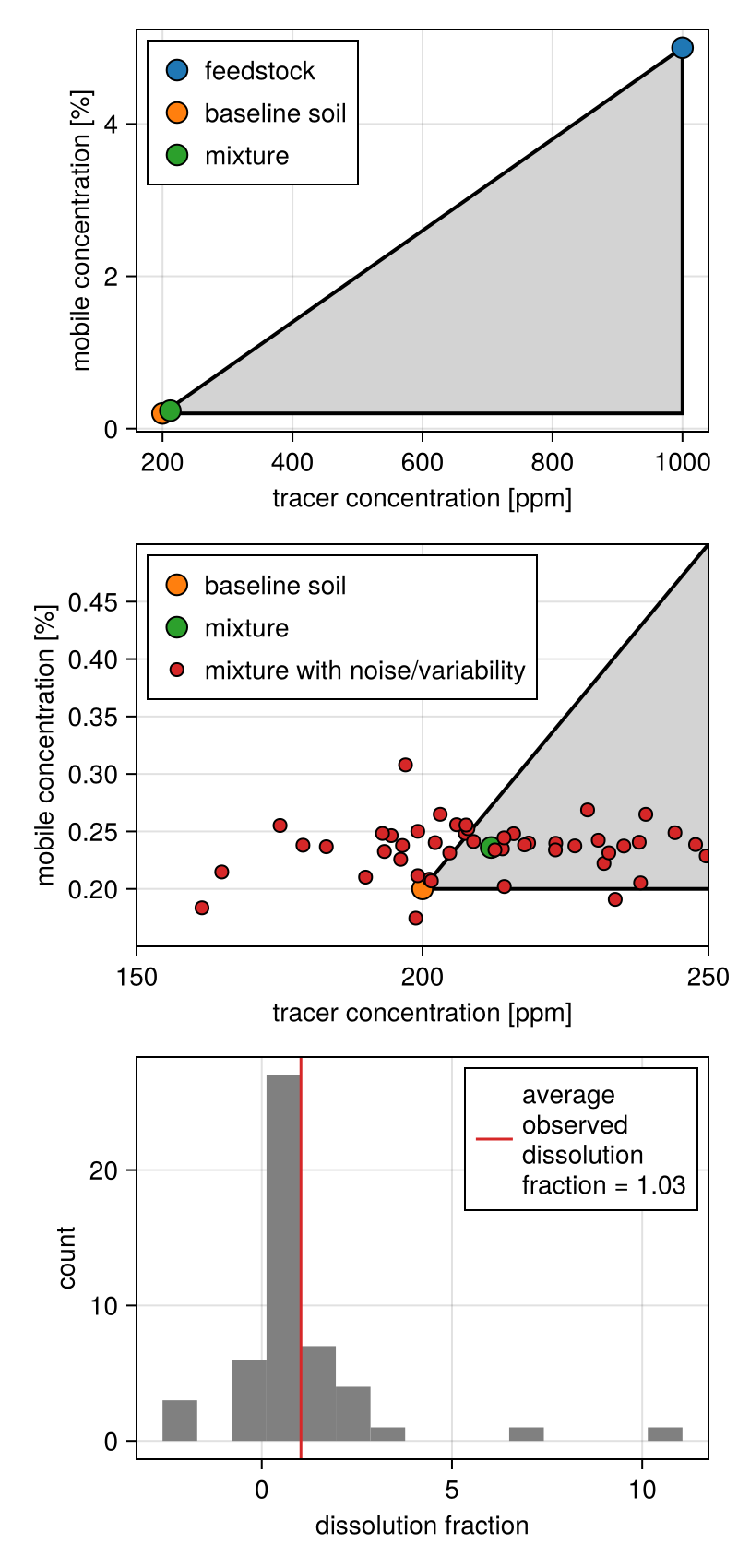}
    \caption{The mixing triangle visualization commonly used for mass-balance tracer methods. See \textcite{reershemius_initial_2023} for a full explanation. \textbf{Top:} Feedstock, baseline soil, and mixed soil coordinates on the tracer-mobile plane. With a realistic mixing fraction like 2 \%, mixed concentrations are very similar to baseline concentrations. The mixed coordinate is very close to the left corner of the triangle. \textbf{Middle:} The same mixing triangle, but zoomed in and with 50 points representing random draws of the mixed concentrations with 10 \% overall relative variability (in red). \textbf{Bottom:} A histogram of the dissolution fractions corresponding to the randomly drawn points in the middle panel and the mean of these dissolution fractions.}
    \label{fig:triangle}
\end{figure}

Figure \ref{fig:triangle} demonstrates how seemingly unexpected dissolution fractions arise from realistic levels of variability in soil concentrations, using the mixing triangle visualization explained in \textcite{reershemius_initial_2023} and elsewhere. We assume a feedstock with mobile element concentration of 5 \% and a tracer concentration of 1000 ppm (by mass). Baseline soil has a mobile element concentration of 0.2 \% and a tracer concentration of 200 ppm. Feedstock is mixed into soil with a mixing fraction of 2 \% and we assume 50 \% of the mobile element has dissolved and leached out of the feedstock.

The top panel of Figure \ref{fig:triangle} shows feedstock and soil end members with the weathered mixture's expected concentration in between. For realistic mixing fractions like 2 \%, mixed concentrations (after spreading and weathering) are very similar to baseline soil. The concentration signals from feedstock addition and weathering are small. The middle panel zooms in on the same baseline and mixed coordinates but includes 50 random points with 10 \% relative variability in both the tracer and mobile concentrations of the mixture (normally distributed). This variability represents everything from natural soil heterogeneity to measurement error and 10 \% is a realistic (arguably optimistic) value.

The variability causes sampled points to stray significantly outside the mixing triangle, producing dissolution fractions that would naively be considered ``unphysical" but are straightforward results of common heterogeneity and noise. The lowest panel shows a histogram of the dissolution fractions corresponding to the sampled points. For this example (50 points), almost half of the points are outside the triangle, some dissolution fractions are significantly outside the [0,1] interval, and the average dissolution itself is also outside the expected interval. These unexpectedly large values are a consequence of the ratios in Equation \ref{eq:dissolutionfrac}. If the denominator $w_t - s_t$ is near small or near zero because of variability and noise, the dissolution fraction will be unexpectedly large. Outlier dissolution fractions can cause the \emph{average} dissolution fraction to be outside the expected interval without a more careful approach.

Moving on from the dissolution fraction, another important issue is that the mixing algebra in section \ref{subsec:tracers} assumes conservation of mass. This directly contradicts the expectation that mobile elements like calcium and magnesium \textit{leave the feedstock-soil mixture} during weathering, which is why $\alpha_t$ and $\alpha_m$ are not equal in the first place. Other bulk elements, like silicon and oxygen, may also leave the mixture as weathering proceeds. The bias introduced by this violation can theoretically be corrected for, as \textcite{reershemius_initial_2023} explore, but correction requires an accurate understanding of the total change in feedstock mass, including all elements, which is difficult and introduces more uncertainty, but could be worth pursuing. The magnitude of the bias depends mainly on how enriched a tracer element is in feedstock compared to soil. More enrichment is always better and further analysis of this issue indicates that tracers should be roughly an order of magnitude more concentrated in feedstock than baseline soil for tracer methods to avoid bias.

Further analysis also shows that dilutive tracers, where the concentration of a tracer element is lower in feedstock than in baseline soil, always introduce unacceptable levels of bias when using the standard tracer approach shown in Figure \ref{fig:prespread_sampling} (also see related criticism by \textcite{reershemius2024error}). Dilutive tracers rely on concentration changes induced by the addition of \emph{bulk feedstock mass}, which is generally a very small signal that is severely biased by the subsequent loss of that mass as weathering proceeds. As such, dilutive tracers should never be used when only pre-spreading and post-weathering samples are available because they don't function as tracers in this context. 

Another important point concerns the use of multiple tracer elements simultaneously, to improve estimates of $\alpha_t$ and ultimately CDR. This is a good idea if multiple immobile elements are enriched enough in feedstock to be well separated from baseline soil concentrations, but care must be taken. \textcite{kantola_improved_2023} attempted a multiple-tracer mass-balance calculation using REEs, with an approach closely related to ``tracer differencing" (Section \ref{subsubsec:disdif}). They used zero-intercept linear regression to estimate the mixing fraction $\alpha_t$, then computed cation concentration changes due to weathering and finally converted those concentration changes to CDR using the density and depth.

\begin{figure}
    \centering
    \includegraphics[width=\linewidth]{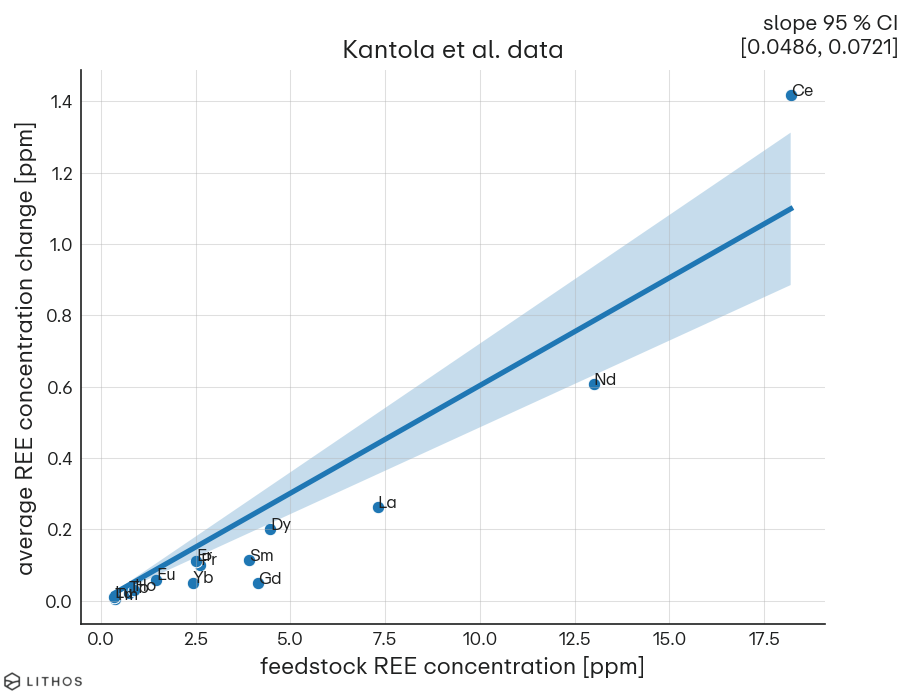}
    \caption{A representation of the mixing regression used in \textcite{kantola_improved_2023}, where the slope of this zero-intercept linear regression is used as an estimate for $\alpha_t$. Measurements for Ce, Nd, and La dominate the regression because the intercept is assumed to be zero and most REEs have much lower concentrations. Low-concentration REEs have almost no effect on the regression, in this case, and are effectively ignored.}
    \label{fig:kantola_regression}
\end{figure}

However, among other very serious issues \cite{reershemius2024error}, \textcite{kantola_improved_2023} do not appropriately scale tracer element concentrations before regression. In their observations, some REE concentrations are naturally much lower than others. Holmium, for example, is much less abundant in their materials than cerium. Consequently, low-concentration elements have almost no weight in the regression. Figure \ref{fig:kantola_regression} shows the data and the resulting zero-intercept regression with a 95 \% confidence interval. The result depends almost entirely on the values for Ce, Nd, and La, effectively ignoring most others. A better approach, instead of regression, is to incorporate multiple tracers into a unified probabilistic model or average appropriately across mixing fraction estimates for each tracer (avoiding differences in scales).

\begin{figure}
    \centering
    \includegraphics[width=\linewidth]{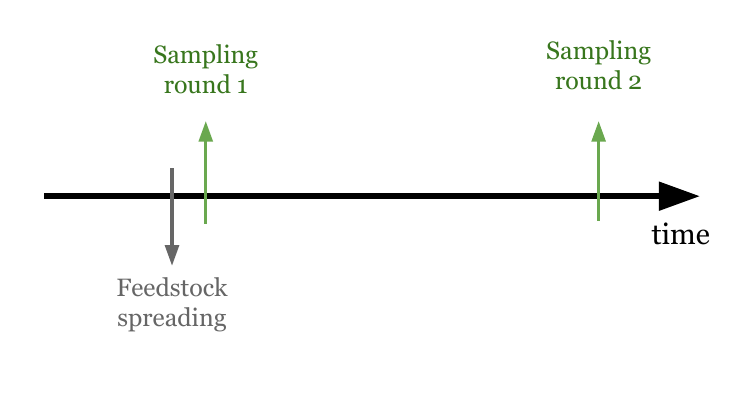}
    \caption{A simple timeline showing sampling rounds after feedstock application and after some period of weathering.}
    \label{fig:postspread_sampling}
\end{figure}

Finally, it's not clear why pre-spreading samples (Figure \ref{fig:prespread_sampling}) have seemingly always been viewed as preferable to post-spreading samples (Figure \ref{fig:postspread_sampling}) when using tracer methods. In the commercial setting, there may be operational reasons to compromise and choose one or the other. However, in controlled trials, the experimenter has much more flexibility. In the post-spreading approach, tracers can be used as a control on changes in density and volume, much like the calculations made in early pedogenesis studies \cite{chadwick_black_1990, brimhall_quantitative_1991}.

Although it makes statistical analysis more involved, it would be even better to sample before \emph{and} after spreading (Figure \ref{fig:both_sampling}), in addition to after weathering. Section \ref{subsec:demo_model} gives an example of one way to approach CDR quantification with samples before and after spreading, using Bayesian inference.

\subsection{Cation Stocks}
\label{subsec:cationstocks}

A practical weakness of tracer methods is that feedstock must be significantly enriched in at least one tracer element compared to the baseline soil (generally by an order of magnitude), and this is not always the case. Without a sufficiently high tracer concentration in feedstock, tracers provide a weak mixing signal and CDR results are vulnerable to bias due to mass loss from weathering. An alternative approach is to ignore tracer elements and estimate changes in cation stocks (total mass) over time. This can be done for two points in time or, if measurements are taken more than two times after spreading, by estimating or inferring the parameters of a time-dependent weathering curve.

Focusing only on cation stocks is conceptually simple and statistically straightforward compared to tracer methods. It's also closely related to soil organic carbon (SOC) practices, where changes in soil carbon stock are monitored over time, and much of the SOC literature is directly analogous to tracking cation stocks. In ERW, however, we are generally accounting for multiple cations simultaneously and we have a strong prior understanding of whether cation concentrations should increase or decrease over time, both of which can be statistical advantages. From that point of view, cation stock monitoring in ERW is like an easier variant of SOC monitoring.

At a minimum, to estimate changes in cation stocks due to weathering, samples are taken soon after spreading and then after weathering has progressed (Figure \ref{fig:postspread_sampling}). Samples are analyzed for concentrations of major cations, along with bulk density, and the difference in each cation stock is used to estimate CDR. There are three main advantages to this approach.
\begin{enumerate}
    \item Changes in cation stocks are what we want to know. Tracer measurements are a means to that end but are not fundamentally necessary.
    \item Feedstock material is selected specifically because of very high concentrations of base cations, typically Ca and Mg. This makes changes in these concentrations after feedstock application, and subsequent weathering signals, large and relatively easy to observe. Unlike tracer elements, there is no scenario in which a feedstock has low concentrations of relevant cations because that would make it unsuitable for ERW and it would not be deployed.
    \item Major cations can be cheaper to measure than tracer elements, an important factor for the scalability of MRV.
\end{enumerate}

There are also some disadvantages to cation stock monitoring. In particular, sample depth \emph{must} be consistent. For example, imagine feedstock is only mixed into the top 5 cm of soil and the first round of samples is 5 cm deep. If the next round of samples is 10 cm deep, cation concentrations will appear to decrease even if no weathering occurred, just because the amount of background soil in each sample has roughly doubled. However, all of these disadvantages can be addressed by controlled and precise sampling operations.

\subsection{Signal Size}

The sections above discuss mass-balance MRV in the abstract. Before discussing simulation and statistics, it's useful to consider some simple, concrete examples of how feedstock application and weathering influence soil concentrations. Concentration changes in soil are the quantities of primary interest for mass-balance methods and having some intuition for the expected magnitudes of these changes is useful when designing simulations and discussing statistical methods.

Assume that a hypothetical feedstock has 6 \% calcium by mass. Before feedstock is spread, the baseline soil has 0.5 \% calcium by mass (5000 ppm). These are representative numbers, but real conditions vary. We spread 10 short tons of feedstock per acre on the soil, or about 2.25 kg/m$^2$. Soil samples are taken after feedstock spreading to 15 cm depth and the soil itself has a bulk density of 1000 kg/m$^3$.

In this scenario, the mixing fraction $\alpha$ (the mass of feedstock per unit mass of soil-feedstock mixture) is 1.48 \%. The calcium concentration of the mixed feedstock and soil is about 800 ppm higher than the baseline soil before any weathering occurs. This 800 ppm change is roughly the size of the ``signal" that must be observed to quantify concentration changes relevant to this hypothetical scenario. That is, to detect the addition of feedstock at all, even right after spreading, we need enough statistical precision to detect an average 800 ppm change between baseline and post-spreading samples (for this scenario).

To resolve weathering over time, we need even more precision. For example, to observe 50 \% loss of calcium from the feedstock would require accurate resolution of, roughly, 400 ppm changes in soil concentration. To observe only a 10 \% loss requires accurate estimation of concentration changes around 80 ppm in magnitude (depending on the method used). If percentages are more intuitive, the 800 ppm calcium enrichment after spreading feedstock constitutes a 16 \% enrichment and the 80 ppm weathering signal is a 1.6 \% change from baseline.

A reasonable generalization is that mixing fractions between 0.5 and 5 \% are typical and concentration signals of around 100-1000 ppm are the observational targets for mobile cations in mass-balance MRV. However, conditions and statistical goals vary, and the strength of the signal is only meaningful in comparison to the size of background variability and noise. 

Tracer elements (Section \ref{subsec:tracers}), however, may not be as highly enriched in feedstock as mobile elements, compared to soil. If a tracer element is present in feedstock at 200 ppm and in baseline soil at 100 ppm, the concentration change immediately after spreading is about 1.5 ppm (1.5 \% enrichment) in this hypothetical scenario. Whatever the sampling and analysis plan is in this case, it will have to accurately resolve changes in this element around 1.5 \%. With many sources of natural variability in soil and limited measurement precision, this may or may not be practical.

\subsection{Variability \& Noise}
\label{sec:variability}

Mass-balance MRV methods for ERW are primarily focused on changes in elemental concentrations over time, in soil. Sometimes we also need more information like sampling depth and bulk density. There are many sources of variability challenging our ability to precisely measure concentration changes and CDR in ERW. They include
\begin{itemize}
    \item Naturally heterogeneous soil concentrations. Most soil is far from a homogeneous soup of elements, but has variable concentrations of cations and tracer elements on a range of spatial scales, in addition to variable bulk density and other characteristics like slope and drainage.
    \item Nonuniform spreading of feedstock. Feedstock is inevitably applied with some level of spatial variability and never as a perfectly uniform sheet over treatment soil.
    \item Treatment area. Although generally well-controlled, the exact boundaries of amended soil can be uncertain.
    \item Feedstock composition. Although generally consistent, the concentrations of elements in the applied feedstock can and do vary.
    \item Sample depth. Because feedstock is not uniformly mixed into a layer of soil with known depth, the sampling depth does influence sampled concentrations.
    \item Real variability in the weathering rate. Nothing prevents the true dissolution rate of feedstock cations from varying, even in a single field. To some extent, this is expected.
\end{itemize}

All of this variability in the field precedes a second layer of noise in the laboratory, where sample handling, preparation, and analysis introduce more imprecision. Laboratory practices and analytical techniques vary and laboratory quality control is a vast topic, but some of the major sources of noise and error relevant to mass-balance are:
\begin{itemize}
    \item Sample splitting. Typically, a mass of soil is dried and pulverized, then a small portion is split from the bulk sample for further preparation and analysis. The splitting process can introduce errors if the material is segregated and poorly homogenized. This is especially true if sample compositing and splitting take place in the field, where homogenization is hard to achieve.
    \item Sample preparation. There are a host of possible errors introduced by further sample preparation. Examples are imprecise weighing, incomplete digestion, contamination, and confusing one sample for another.
    \item Analysis. All instruments are subject to some level of noise.
\end{itemize}
Mass-balance MRV methods must be able to see weathering signals through all these sources of variability, error, and noise by averaging over them.

\section{Planning Experiments}
\label{sec:planningexperiments}

There are many possible ways to do mass-balance MRV for ERW. Experimenters and practitioners have to decide when to pull soil samples, where to collect them, how many to take, how many individual cores to composite for each sample (if any), and which quantities to measure when the samples are in hand. They have to decide whether tracer elements should be used and which elements are tracers. They have to decide how to handle control/treatment groups. Ultimately, initial CDR has to be computed from some set of observations, somehow.

Ideally, these decisions are made based on
\begin{enumerate}
    \item an informed understanding of variability and noise in the field and laboratory
    \item the desired statistical precision or confidence of the result
    \item clear, pre-documented understanding of exactly what data will be collected and what specific analyses will be performed with the data: a stated plan and an estimand
\end{enumerate}
However, there has been a general lack of rigorous planning in ERW research. Here are just a few examples from mass-balance studies.

\textcite{reershemius_initial_2023} performed replicated mesocosm experiments to measure weathering rates in a controlled setting, but only took one baseline soil measurement for the whole group of replicates, making it impossible to account for baseline variability across replicates and measurement error in the individual baseline. \textcite{kantola_improved_2023} used a set of REEs as tracer elements for CDR quantification, but almost all of the candidate tracers were insufficiently enriched in feedstock and, in some cases, less abundant in feedstock than in soil, making them unsuitable. These authors also group and average their data in surprising ways without persuasive justification (see also comments in Section \ref{subsubsec:comcav} and criticism by \textcite{reershemius2024error}). Using the same samples from the same trial as in \cite{kantola_improved_2023}, \textcite{beerling_enhanced_2024} split samples into two sections by depth and state that almost half of the applied feedstock was mixed into the lower layer. Analysis of these deeper samples is then omitted, without explanation, and the authors extrapolate results from the upper layer to the lower one, claiming that this state of ignorance is the ``conservative" choice.

Experiments and field trials are difficult and sometimes messy, but there is a need to more rigorously inform and justify choices like sample size, sample timing, and sample depth. The experimental plan and analysis plan should be formulated and tested \emph{before} the experiment, not after, to discover and address problems that are impossible to correct later on (like too few baseline samples and bad tracers). Decisions in data analysis, like which groups of samples to use and how to average across groups, should be made during the planning process, not after reviewing final data. Researchers should avoid calculating results with different slices or groups of data, presenting the individual result that looks best according to preconceptions, and coming up with post hoc explanations for their choices, as this straightforwardly introduces bias.

There is also a need to validate statistical methods used to estimate CDR. This can be difficult in complex media like soil and with complex sampling plans. However, demonstrating that statistical methods and their implementations are \emph{even capable} of reliably producing the correct answer is critical to continue building an empirical understanding of weathering rates for ERW.


The section below explains how to simulate and understand hypothetical deployments or experiments \emph{before} carrying them out, to minimize the risk of errors and omissions, then presents a publicly available software package for application to mass-balance methods in ERW. Later on, we show a mildly simplified example initial CDR quantification using Bayesian inference.

\section{Simulation}
\label{sec:simulation}

A robust and flexible way to understand the combined effects of experimental design and variability/noise is to simulate the data generating process \cite{mcelreath2018statistical}. In this approach, we assume an experimental design and prescribe probability distributions for relevant physical and chemical parameters. Then we simulate from these assumptions. We draw random realizations of our experimental plan and physical parameters jointly, generating synthetic datasets with randomness representative of real conditions.

In the context of mass balance for ERW, we draw realizations of soil samples in space and time, along with realizations of feedstock and soil parameters, producing complete sets of simulated geochemical data. The hypothetical datasets generated by this Monte Carlo approach are representative of real MRV datasets produced by field trials or commercial deployments and can be used in a number of important ways.
\begin{enumerate}
    \item Understanding the system. Modeling different experimental designs and physical conditions forces us to think about which parameters are relevant, how they vary, and why. It can also expose overlooked sources of variability or interactions between parameters.
    \item Uncertainty analysis. Simulated datasets directly address questions about how variability in physical parameters contributes to uncertainty in the final statistical result. We can simulate datasets with different levels of variability, feed them into a statistical method of choice, and see how uncertainty in our results corresponds to variability in the parameters.
    \item Statistical validation. Simulated data is produced by a forward model for which \emph{we know the correct answer unambiguously}. For mass balance, we know how much feedstock dissolved and leached for all simulated datasets because we prescribe the model parameters. To validate statistical methods, we feed simulated datasets into our methods and check that they can recover the correct answer, on average. We can also evaluate whether a statistical method provides an accurate representation of uncertainty. For example, if we are using 90 \% confidence intervals to evaluate uncertainty, we can check directly that our intervals contain the correct (known) answer for about 90 \% of simulated datasets.
    \item Power analysis. Synthetic data can be used to choose an appropriate number of samples based on a desired level of statistical precision. To do so, we can make cautious assumptions about variability and noise, simulate datasets with different sample sizes, and see how many samples are required to achieve the desired level of precision for the statistical method we plan to use for real data. Although not perfect, this approach is far better than the common strategy of guessing.
\end{enumerate}

All of these tasks involve the same general process. Simulated datasets are the \emph{joint} products of sampling choices and natural variability/noise and can be analyzed as if they were real datasets to answer questions about how sampling choices and variability affect the analysis. This includes complications like control samples, many rounds of sampling, laboratory error, composite sampling, and so on. Simulations can guide sampling choices and statistical practice.

\subsection{\texttt{Monty}}

We present a software package for performing the simulations described above in the context of mass-balance MRV for ERW, called \texttt{Monty} (a reference to the Monte Carlo approach). The package is written in the Julia programming language \cite{Bezanson_Julia_A_fresh_2017}. The central element of the package is a deterministic mixing model that describes how the elemental concentrations and mass of an individual soil core are determined by relevant physical and chemical parameters. On top of this model is a set of functions for defining sampling plans and the variability of mixing parameters like baseline soil concentrations, application rates, etc.

First, we describe this mixing model in detail, before highlighting some other useful aspects of the package. Other details are left to the \href{https://lithoscarbon.github.io/Monty.jl/}{online documentation}, where there are also examples showing how to use the primary components of the package. In addition to the documentation, the package is thoroughly \href{https://github.com/LithosCarbon/Monty.jl/actions/workflows/tests.yml}{tested} and \href{https://github.com/LithosCarbon/Monty.jl/actions/workflows/benchmarks.yml}{profiled}.

\subsubsection{Deterministic Mixing Model}
\label{subsubsec:mixing}

To simulate the mixing of baseline soil and weathered feedstock in an individual soil core, the model assumes a vertically homogenous layer of baseline soil. This soil is characterized by the bulk density ($\rho_s$) and concentrations of key elements in the soil ($c_s$). Any number of elements can be included. Feedstock is added to this layer according to an application rate ($Q$) and mixed vertically into the soil according to a mixing profile ($\Gamma$) represented by any arbitrary, non-negative probability distribution so that the integral of the profile over infinite depth is always equal to one (original feedstock mass is not lost). The feedstock has prescribed elemental concentrations ($c_f$) and can have a nonzero bulk density ($\rho_f$) so that it occupies vertical space in the soil. Over time, elements can be dissolved and leached out of the feedstock, with each element assigned a monotonically increasing (or flat) loss fraction ($l$) for all points in time. In addition to the loss of specific elements, some of the bulk feedstock mass is also lost over time, represented by a total mass loss fraction ($L$). A soil core is taken from this mixture with some depth ($d$) and cross-sectional area ($a$).

The elemental concentrations of the soil core are determined by adding the masses of each element in soil and feedstock then dividing by the total mass. The fraction of original feedstock present in the soil core ($\gamma$) is determined by the feedstock mixing profile and the depth of the core,
\begin{equation}
    \gamma = \int_0^d \Gamma(x) dx \, .
\end{equation}
The total mass of feedstock in the core is
\begin{equation}
    M_f = Q \gamma (1 - L) \, ,
\end{equation}
which accounts for the feedstock mass loss fraction, $L$. The vertical space occupied by the sampled feedstock is
\begin{equation}
    h = M_F / \rho_f \, .
\end{equation}
The mass of each element of interest in the core located in feedstock ($m_{f,i}$), per unit area, is
\begin{equation}
    m_{f,i} = Q \gamma (1 - l_i) c_{f,i}
\end{equation}
The total mass of soil in the core, per unit area, accounting for space occupied by feedstock, is
\begin{equation}
    M_s = \rho_s (d - h) \, ,
\end{equation}
and the mass of each element of interest in the core from the baseline soil, also per unit area, is
\begin{equation}
    m_{s,i} = \rho_s (d - h) c_{s,i} \, .
\end{equation}
The total mass of the core is
\begin{equation}
    M = M_f + M_s =  Q \gamma (1 - L) + \rho_s (d - h)
\end{equation}
and the concentrations of each element in the soil core (per unit area) are computed by adding up each element's mass and dividing by the total mass,
\begin{align}
    c_i &= \left( m_{f,i} + m_{s,i} \right) / M \\[1ex]
    &= \frac{Q \gamma (1 - l_i) c_{f,i} + \rho_s (d - h) c_{s,i}}{Q \gamma (1 - L) + \rho_s (d - h)} \, .
\end{align}
This is a physically consistent, although moderately simplified, representation of the concentration and mass of an individual soil core. The model assumes feedstock is mixed into a vertically homogeneous layer of soil and feedstock loses cations to weathering at prescribed, depth-independent rates. Loss rates can be different for each element. Table \ref{tab:mixingparams} organizes all mixing model parameters and their SI units. 

A key aspect of this simple model is the prescribed loss fractions $l_i$. This is how the correct answer, with respect to weathering rates, is known when synthetic data is eventually generated. We prescribe the functions $l_i(t)$ and know exactly what the loss rates are.

\begin{table*}
    \centering
    \small
    \begin{tabular}{r|c|p{0.6\textwidth}}
        $\Gamma(d)$ & kg/kg & feedstock mixing profile, a non-negative probability distribution describing the vertical distribution of feedstock mass mixed into the soil \\[1ex]
        $d$ & m & sample depth \\[1ex]
        $\gamma(d)$ & kg/kg & the fraction of original feedstock mass mixed into the soil collected by a soil core, computed via the CDF of $\Gamma$ and the sample depth $d$\\[1ex]
        $a$ & m$^2$ & soil core cross-sectional area \\[1ex]
        $Q$ & kg/m$^2$ & feedstock application rate \\[1ex]
        $\rho_f$ & kg/m$^3$ & feedstock bulk density as it exists in the soil-feedstock mixture \\[1ex]
        $c_{f,i}$ & kg/kg & feedstock concentration of element $i$ \\[1ex]
        $\rho_s$ & kg/m$^3$ & soil bulk density \\[1ex]
        $c_{s,i}$ & kg/kg & soil concentration of element $i$ \\[1ex]
        $l_i$ & kg/kg & loss fraction of element $i$ from feedstock \\[1ex]
        L & kg/kg & loss fraction of all feedstock mass 
    \end{tabular}
    \captionsetup{width=0.7\textwidth}
    \caption{Parameters (inputs) used in the \texttt{Monty} mixing model to compute the exact elemental concentrations and mass for a single soil core, along with their SI units and brief descriptions.}
    \label{tab:mixingparams}
\end{table*}

\subsubsection{Composite Sampling}

It's common practice in agricultural and soil sampling to physically combine individual soil cores into a single sample, which is then homogenized and analyzed. This practice averages over the variability of individual cores, but increases the importance of laboratory noise by decreasing the number of individual measurements. On this topic, much can be learned from the SOC literature (see \textcite{spertus_optimal_2021} and references therein).

\texttt{Monty} makes it very straightforward to model soil core compositing. From the mixing model in Section \ref{subsubsec:mixing}, each soil core is represented by the total mass of soil and the concentration of each element of interest. Compositing one or more cores means taking the mass-weighted average of each element's concentration. In \texttt{Monty} this is done using the addition operator/function (\texttt{+}). To composite samples, they are simply added together, and any number of cores can be composited by summing them (the \texttt{sum} function).

The package also implements a number of common compositing ``stencils," which are spatial patterns of the individual soil cores that are combined into a single composite sample. Although some theoretical statistical work is based on compositing random groups of cores for an entire field, this is almost never done in practice. Much more commonly, some number of cores are taken at individual sample locations and composited immediately. This practice averages over short-distance variability. An example of automatic stenciling and compositing is shown later in Section \ref{subsec:demo_model}.

\subsubsection{Spatial Covariance}

A primary area of focus for mass-balance MRV is the natural spatial variability of elemental concentrations in soil, although it has been overlooked in some prior work. For example, one line of research \cite{reershemius_new_2023} posited that 1 \% analytical precision (relative standard deviation of measurement error) is sufficient to resolve mass-balance weathering in an \emph{individual} soil sample, grossly overlooking the impact of variance in natural soil (among other sources of variability) and generally neglecting any serious treatment of uncertainty. In fact, although conditions vary and laboratory precision is important, sources of variability outside of the lab usually dominate the overall noise. As such, understanding and planning adequately for baseline variability in experiments, field trials, and commercial deployments is absolutely critical.

\texttt{Monty} addresses spatial variability explicitly, allowing baseline soil concentrations to be sampled from arbitrary multivariate distributions. This includes Gaussian Processes---multivariate normal distributions with the covariance matrix structured by the distances between points---a natural choice for inducing spatial autocorrelation in soil concentrations.

The spatial structure of a field's properties can be conveniently defined by a theoretical variogram. These are common functions in geostatistical applications that describe the similarity between points based on their distance and can be used to define the covariance matrix of a multivariate normal distribution representing that similarity. Importantly, these functions allow for variability even at very close (nearly zero) distances. Variability on short distances is commonly observed in real soil and it determines the utility of compositing strategies.

\subsection{Single Core Sensitivity}
\label{subsec:sensitivity}

Here we examine the sensitivity of a single soil core's elemental concentration, using \texttt{Monty}'s mixing model (Section \ref{subsubsec:mixing}). Before complex simulations with many sources of variability and noise, we build some intuition for what parameters fundamentally influence the concentration of any given sample. The balance of this influence indicates which parameters are important, as sources of variability in the field, for mass-balance MRV. This kind of analysis is relatively simple but illuminating.

\subsubsection{Local Sensitivity}
\label{subsubsec:localsens}

First, the most basic sensitivity analysis. We assume representative values for the mixing parameters and look at the model's gradient with respect to those parameters, using convenient automatic differentiation \cite{RevelsLubinPapamarkou2016} tools. This shows us the first-order effect of each parameter on the elemental concentration of a hypothetical soil core.

For this example, we examine only one element's concentration, using soil and feedstock concentrations representative of a base cation like calcium mixed into agricultural soil. We ignore the core's cross-sectional area ($a$) because it only influences the core's total mass, not its concentration. The parameters are
\begin{itemize}[itemsep=1pt,parsep=1pt]
    \item $\gamma = 0.9$ kg/kg
    \item $d = 10$ cm
    \item $Q = 3$ kg/m$^2$ (about 13 short ton/acre)
    \item $\rho_f = 3000$ kg/m$^3$
    \item $c_f = 0.05$ kg/kg (50,000 ppm)
    \item $\rho_s = 1000$ kg/m$^3$
    \item $c_s = 0.003$ kg/kg (3,000 ppm)
    \item $l = 0.5$ kg/kg
    \item $L = 0.5$ kg/kg
\end{itemize}

\begin{figure}
    \centering
    \includegraphics[width=0.45\textwidth]{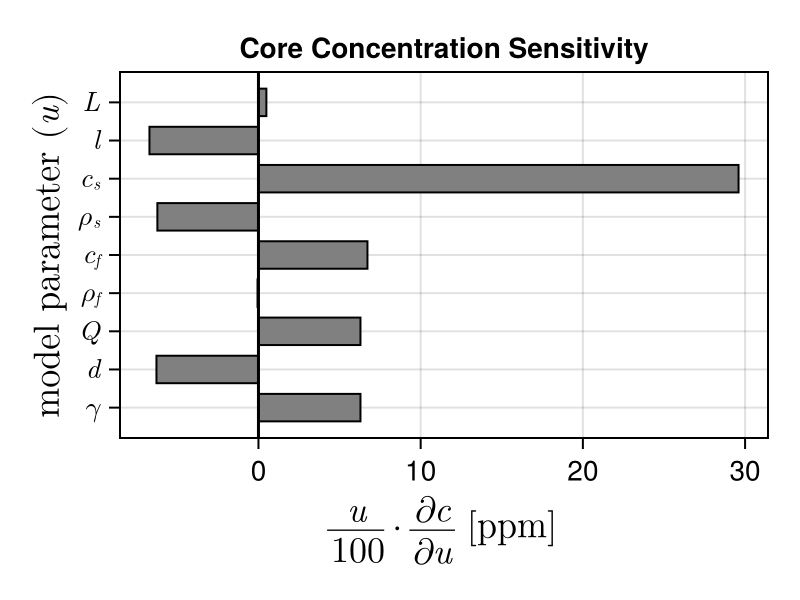}
    \caption{The relative gradient of the mixing model for one set of inputs, showing how much a soil core's concentration changes due to +1 \% changes in each input parameter, in units of ppm.}
    \label{fig:gradsensitivity}
\end{figure}

Figure \ref{fig:gradsensitivity} shows how much the elemental concentration of a soil core changes when the input parameters are changed. It shows the partial derivative for each parameter multiplied by 1 \% of its input value. So, each bar shows how much the soil core's concentration changes for a 1 \% change in the parameter, in units of ppm for convenience.

Some model parameters are almost negligible in this scenario. Changes in the feedstock mass loss ($L$) and the feedstock bulk density ($\rho_f$) have almost no impact on the soil core's concentration. This is straightforward to understand. The feedstock mass is only a small portion of the overall core's mass, about 1.5 \%. The rest of the core's mass comes from baseline soil. As such, changes in the total mass of the feedstock ($L$), independent of changes in the element of interest, barely matter. For the same reason, the vertical space occupied by the feedstock is minimal, so it displaces almost no background soil and its bulk density ($\rho_f$) is immaterial.

Core concentration is sensitive to all the other parameters with this set of input values. The element loss fraction ($l$) predictably causes a decrease in the core concentration. Increasing soil bulk density and sample depth also decreases core concentration by increasing the soil mass in the core and diluting the feedstock's contribution.

The baseline soil concentration is the most important mixing parameter. This is explained by the fact that almost all of the mass in the soil core is derived from baseline soil, not feedstock. The feedstock mixing fraction is only about 1.5 \% in this case and will generally be no larger than a few percent under realistic conditions, so changes in soil concentration almost always translate to nearly equal changes in core concentration. Put another way, if $c$ represents the mixed core's concentration, $\partial c / \partial c_s \approx 1$.

This is only one set of input parameters and the sensitivities shown here are meaningfully different from other choices. For example, the application rate ($Q$) has been very high in some experiments and field trials. In that scenario, changes in all feedstock parameters will have more influence on core concentration. It's also important to note that a 1 \% change in each input parameter is an arbitrary choice. What ultimately matters is how much each parameter varies in the experiment or field. For example, baseline soil concentrations could vary by 30 \% spatially and sample depth by much more than 1 \% (which is only a millimeter for a 10 cm core). The balance of each parameter's variability determines its importance for core concentration variability.

\subsubsection{Global Sensitivity}

Another way to quantify the importance of mixing parameters is with global sensitivity analysis (GSA). Here we use the Sobol method \cite{dixit2022globalsensitivity}, which decomposes the variance of model output into fractions that that are attributable to individual parameters or groups of parameters. Importantly, the method samples from a range of values for each input.

As discussed above, the correspondence between the concentration of a soil core and that of the baseline soil is very strong unless there is an extremely high application rate. For that reason, soil concentration ($c_s$) is left out of this GSA and set to 3,000 ppm. We use one million samples from the model with the following parameter ranges:
\begin{itemize}[itemsep=1pt,parsep=1pt]
    \item $\gamma \in [0.1, 1]$ kg/kg
    \item $d \in [5, 25]$ cm
    \item $Q \in [0, 5.6]$ kg/m$^2$ (up to 25 short ton/acre)
    \item $\rho_f \in [1000, 3000]$ kg/m$^3$
    \item $c_f \in [0.04, 0.1]$ kg/kg (40,000 to 100,000 ppm)
    \item $\rho_s \in [500, 1500]$ kg/m$^3$
    \item $l \in [0, 1]$ kg/kg
    \item $L \in [0, 1]$ kg/kg
\end{itemize}
Note that the results of the GSA \emph{do} depend on these parameter ranges. They represent plausible ranges for most experiments, trials, and deployments.

\begin{figure}
    \centering
    \includegraphics[width=0.45\textwidth]{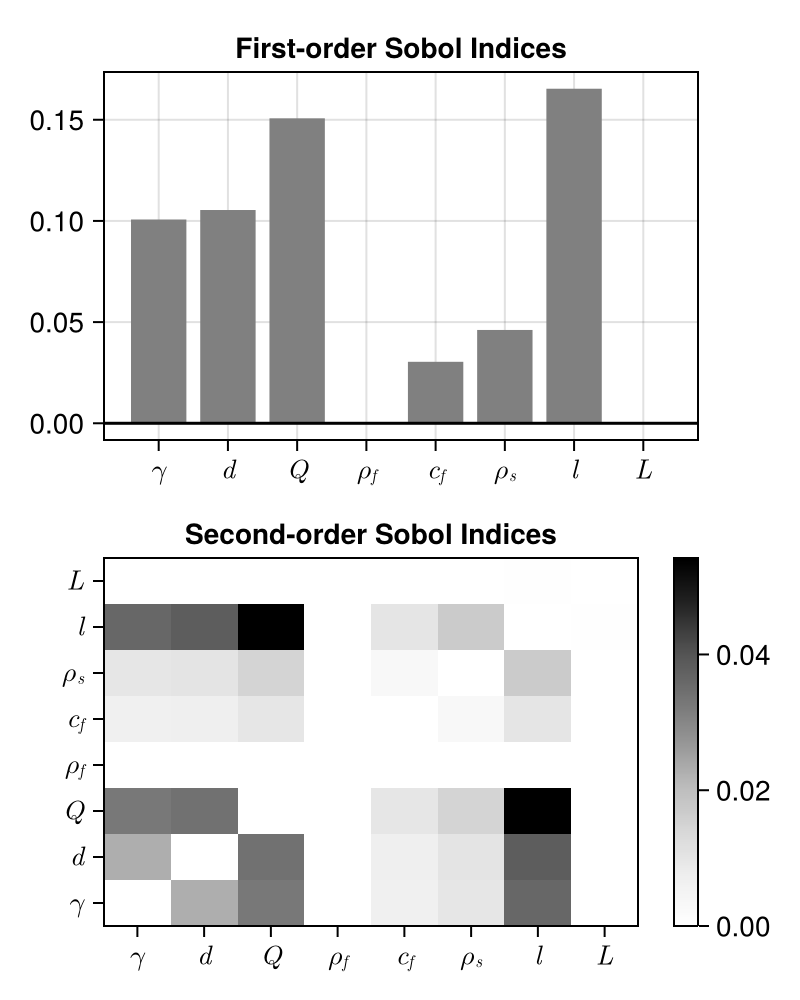}
    \caption{First- and second-order Sobol indices for mixing parameters, ignoring the baseline soil concentration (set to 3,000 ppm).}
    \label{fig:sobol}
\end{figure}

Figure \ref{fig:sobol} shows Sobol indices for these parameters. The upper panel shows first-order indices, which quantify the direct contribution of each parameter to core concentration variance, ignoring contributions from interaction with other parameters. As in the local sensitivity analysis (Section \ref{fig:gradsensitivity}), the feedstock bulk density and total mass loss have very little effect on core concentration. Again, this is because feedstock mass represents a small portion of the core's total mass under plausible conditions.

The application rate ($Q$) and elemental loss fraction ($l$) have the strongest first-order effects on core concentration for the chosen parameter ranges. The sampled feedstock fraction ($\gamma$) and core depth ($d$) are also meaningful. The feedstock concentration ($c_f$) and soil bulk density ($\rho_s$) are less influential, but not negligible.

Second-order Sobol indices indicate that the application rate and loss fraction interact most strongly with other parameters and with each other. These interactions have moderate overall effects on the ultimate core concentration, compared to the first-order effects. Here again, we see no effect of total feedstock mass loss or feedstock bulk density. There are some comparatively weak interactions between soil bulk density, feedstock concentration, and other parameters.

\subsubsection{Sensitivity Summary}

The most important aspect of soil-feedstock mixtures in ERW is that the feedstock mass likely makes up no more than a few percent of the mixture's total mass. Generally, $\alpha \leq 0.05$. The \emph{total mass} of a sample is only barely modified by feedstock application and loss unless the application rate is extremely high and/or sampling is very shallow. This explains the mixture concentration's insensitivity to feedstock bulk density ($\rho_f$) and bulk mass loss ($L$).

Low feedstock mixing fraction also explains the dominant influence of baseline soil concentration on sample concentration for almost any element. It might seem obvious, but because baseline soil comprises nearly all of the mass in any given sample, changes in baseline soil concentrations strongly influence mixed concentrations.

Feedstock addition and feedstock properties do matter, but their effects are usually diluted by significant amounts of background soil mass. The local sensitivity analysis (Section \ref{subsubsec:localsens}) also discusses this. Exceptions include cases with extremely low soil concentrations or extremely high feedstock application.

However, it's also important to remember that these sensitivities are only relevant in the context of variability. If the baseline soil concentration is very homogenous, its impact is attenuated. Similarly, if the application rate is highly spatially variable, its impact is amplified. 

\section{Estimation \& Inference}
\label{sec:estimationinference}

As reviewed in Section \ref{sec:mass-balance}, there are many possible ways to compute CDR from the geochemical data produced for mass-balance MRV methods. These data consist, at minimum, of elemental concentration measurements at two different times during an experiment or deployment, including the major cations and possibly other elements as tracers. In some cases, information like sample depth and soil density is also necessary.

The goal of mass-balance MRV is to use these observations to estimate or infer initial CDR for some domain (experiment, field, group of fields, etc.) with uncertainty. The literature on experimental statistics is vast. We only discuss the most salient topics below. 

\subsection{Estimating CDR}

ERW practitioners estimate average cation concentration changes and their uncertainties as a proxy for initial CDR. Some authors use error propagation methods \cite{kantola_improved_2023, beerling_enhanced_2024}, usually by assuming each component of an expression for CDR has normally distributed uncertainty and propagating that uncertainty through an expression for CDR. This approach can be workable but has limitations for a complex estimand like CDR and with more complex datasets.

First, it often assumes normally distributed errors, which may not be appropriate. Second, as briefly discussed in Section \ref{subsubsec:disdif}, using tracers to infer the mixing fraction $\alpha_i$ requires division by $w_i - s_i$. If the error in this term is large, it might overlap zero and include significant probability mass on negative values, which is problematic. If the propagation is carried out by Monte Carlo sampling, for example, samples will occasionally produce extremely large (positive or negative) values when $w_i - s_i \approx 0$. More generally, as the probability density on zero increases in the denominator ($w_i - s_i$), the ratio becomes increasingly badly behaved and ultimately a Cauchy distribution. Third, error propagation becomes very unwieldy for more complicated datasets and models. For example, it would be difficult to use error propagation in a scenario where several rounds of samples are available over time and the goal is to estimate the parameters of a weathering curve, using all of the available data simultaneously. It's better to take a more unified statistical approach.

Another compelling option for computing average initial CDR with uncertainty is bootstrapping \cite{kulesa_sampling_2015}. Non-parametric bootstrapping assesses the uncertainty of a sample estimate by resampling from the observations with replacement, repeatedly calculating the desired quantity (CDR in this case) from the resampled cases. Importantly, resampled datasets are the \emph{same size} as the original, which can be a point of confusion, and the observations themselves must be independent. Bootstrapping is particularly useful when the quantity of interest is a complex function of the observations, which is usually true for mass-balance methods, especially when using tracers. This flexibility is a major strength and makes bootstrapping a good choice for CDR estimation. However, as with error propagation, resampled estimates can produce pathologically large values when estimating the dissolution fraction (Equation \ref{eq:dissolutionfrac}) and for the same reason.

\subsection{Bayesian Inference}

Bayesian modeling is another way to understand how much initial CDR occurred in an experiment or deployment. It is a deep topic and we recommend \textcite{mcelreath2018statistical}, \textcite{kruschke2014doing}, and \textcite{gelman_bayesian_2020} for those unfamiliar. To our knowledge, Bayesian models have not been used for field trials in ERW or by other commercial suppliers but they have some significant advantages for mass balance MRV and for ERW quantification broadly. We summarize some of these advantages below in the context of ERW.

First, Bayesian models allow for explicit, rigorous incorporation of physical context and constraints---prior understanding. This context can both improve the accuracy of our results for CDR and include sources of uncertainty that are known to exist but difficult to incorporate with other statistical methods. For example, one component of mass-balance methods that is often overlooked is the fraction of applied feedstock mass that is contributed by water adsorbed to the surface of fine particles. This moisture fraction is generally not negligible and directly affects the total amount of CDR that can occur through ERW, but it's never known exactly. Bayesian models can incorporate this uncertainty directly.

Even with quality control practices, the same is also true for other deployment parameters. We generally do not know, with perfect precision, the exact feedstock mass spread on the treatment area, elemental concentrations in feedstock, and so on. These parameters influence CDR and Bayesian models can incorporate their uncertainty naturally. Such models can also enforce physical constraints, like the fact that total CDR must not imply that more feedstock dissolved than was initially spread on the field.

Second, Bayesian models are flexible. They can represent complex processes and incorporate diverse sources of data jointly. For example, they can incorporate any number of tracer elements, any number of cations, and any number of sampling rounds coherently, all at the same time. They can model known physical relationships, like the general expectation that calcium and magnesium dissolution rates are not radically different. They can be structured hierarchically, for example when partially pooling information across groups of experimental replicates or entire deployments. Ultimately, data from every deployment managed by an institution or company could be incorporated into a single unified model.

Third, the result of Bayesian modeling is a posterior distribution, which has a very direct interpretation. This can be an advantage compared to confidence intervals or p-values, which are frequently misinterpreted. Posterior distributions can also be an advantage for subsequent decision-making or further incorporation of uncertainty. The posterior distribution for CDR is just a probability distribution: a description of how probable different values are.

Fourth, after conditioning on data, Bayesian models can be used to predict subsequent observations with clear incorporation of the uncertainty for each parameter in the model. In ERW, for example, properly structured models could learn to predict dissolution rates from correlated associations between observations of soil acidity, cation exchange capacity, porosity, soil type, hydrology, etc. Bayesian models can always be used to both infer and predict. This is not workable with error propagation methods, for example.

At the same time, one of the strengths of Bayesian models---grounding in prior understanding and physical knowledge---can also be a weakness. To be credible, the assumptions made by such models must be fully transparent, justifiable, and open to refinement and criticism. This includes both prior distributions and the model structure itself. It is also critical to remember that the flexibility of these models does not correct for inadequate sample size, bad experimental design, or other errors. Collecting a sufficient amount of high-quality data from well-planned trials is always the priority.

\subsection{Demonstration Model}
\label{subsec:demo_model}

Here we demonstrate the use of a Bayesian model for CDR inference in the mass-balance framework. We use simulated data from \texttt{Monty} because we can dictate the correct answer underlying the simulated data and confirm that the model behaves as expected. For a Bayesian model, we can check that the posterior distribution contracts toward the correct values for each parameter and, with more data, contracts further. In this case, we ignore tracers, simulating and then inferring CDR using only the major cations calcium and magnesium. Tracers (any number of them) can certainly be incorporated into a Bayesian model like the one presented here, but we ignore them for brevity and clarity.

\subsubsection{Hypothetical Deployment}
\label{subsubsec:hypodepo}

\begin{figure}
    \centering
    \includegraphics[width=0.45\textwidth]{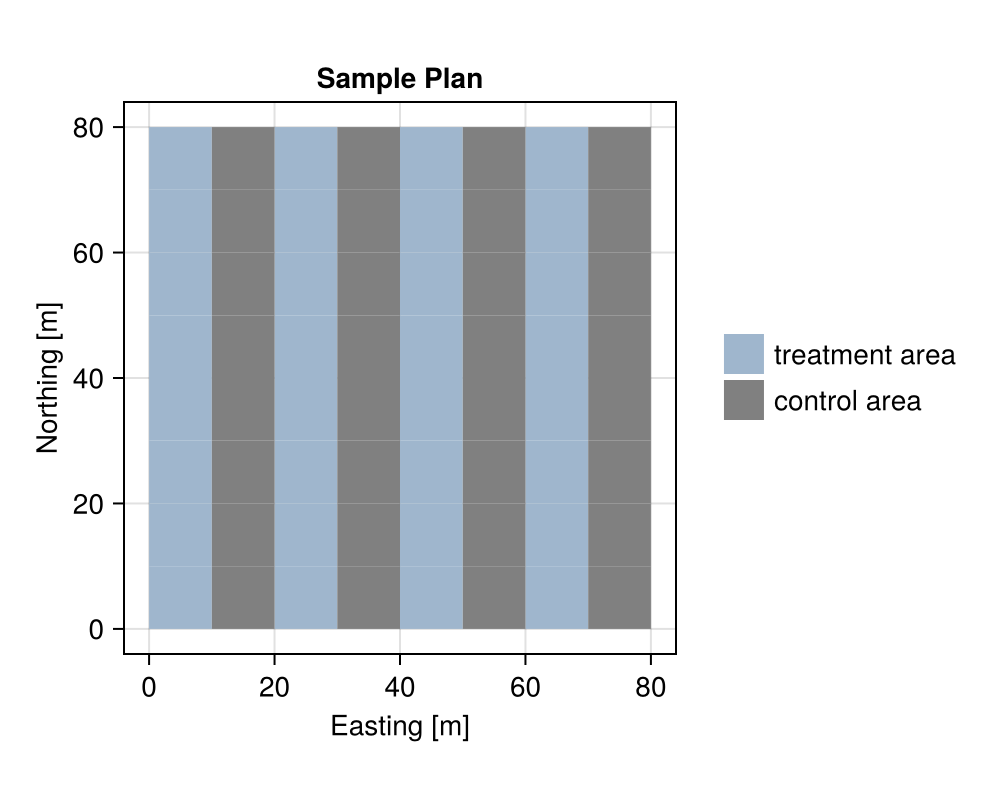}
    \caption{The arrangement of the hypothetical deployment used to generate data for the Bayesian modeling example.}
    \label{fig:plan}
\end{figure}

\begin{figure}
    \centering
    \includegraphics[width=0.45\textwidth]{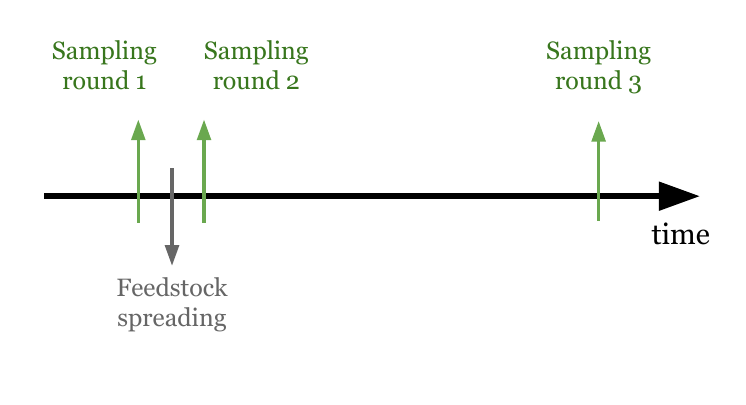}
    \caption{A simple timeline showing sampling rounds before feedstock application, after feedstock application, and after some period of weathering. Three rounds of sampling.}
    \label{fig:both_sampling}
\end{figure}

To produce simulated data, we imagine a hypothetical field trial on a small parcel of land, only 1.58 acres (6400 m$^2$). The parcel is divided into an 8 by 8 grid of cells. Columns of the grid alternate between treatment and control groups. Figure \ref{fig:plan} shows the spatial arrangement of the hypothetical trial. One composite sample is taken from each cell for each of three sampling rounds:
\begin{enumerate}[itemsep=1pt,parsep=1pt]
    \item immediately before spreading
    \item immediately after spreading
    \item one year after spreading
\end{enumerate}
Figure \ref{fig:both_sampling} shows the timeline.

Target sample locations are selected randomly from within each cell, but the target location for each cell is identical for all three rounds of sampling. Random, normally distributed error ($\sigma=0.75$ m) is applied to the realized location of each sample, representing unavoidable inaccuracy in the hypothetical soil sampler's GPS and point-finding ability.

Each sample is a composite made up of five individual soil cores arranged in a circle with a radius of 2 m around the sample location. Random normally distributed error is also applied to each core's location ($\sigma = 10$ cm). Soil core depths vary between 5 cm and 15 cm according to a symmetric triangular distribution.

The whole sampling plan consists of 960 individual soil cores which are composited into 192 samples, then measured. In each sampling round, there are 32 samples in both the treatment and control groups. To be clear, we are not indicating that this sample plan is optimal, but using it to demonstrate simulation and modeling capabilities. Also, just for reference, laboratory analysis of all samples at \$50 per sample would cost \$9600.

\begin{figure*}
    \centering
    \includegraphics[width=\textwidth]{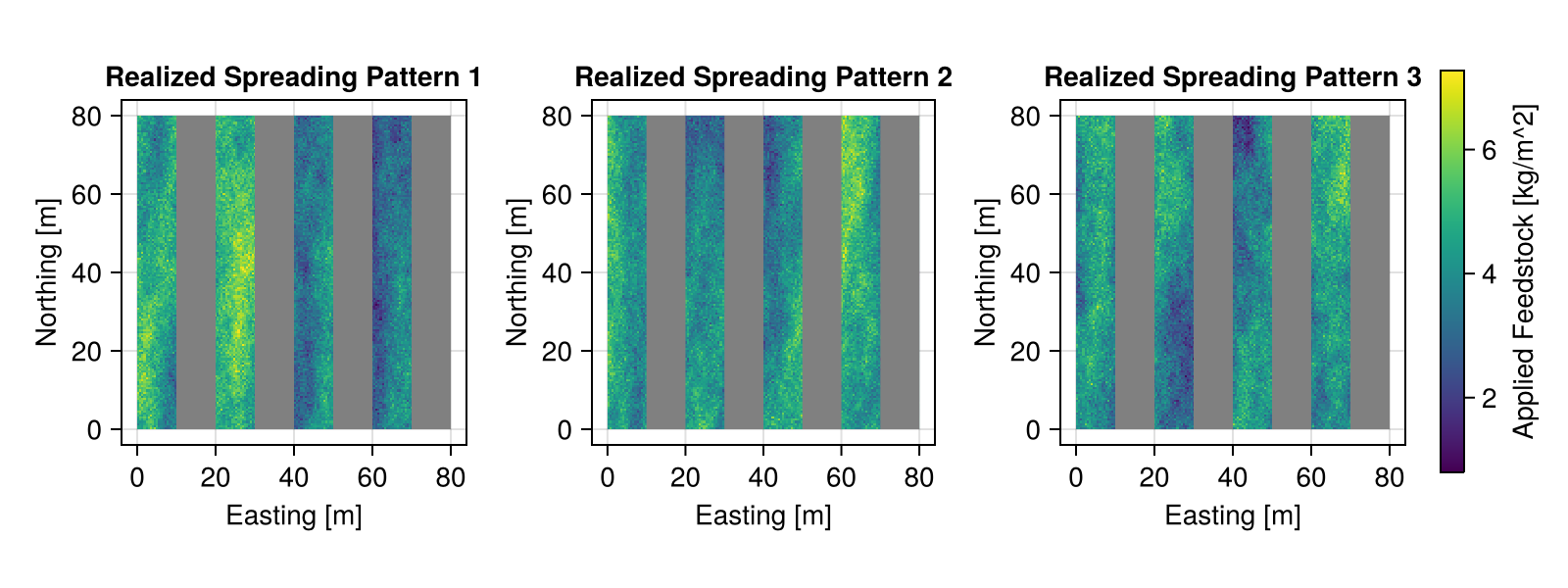}
    \caption{Three realizations of the Gaussian process used to generate application rates at sample points. A ``realization" is a random draw from the multivariate Gaussian distribution produced by the Gaussian process. In these cases, the Gaussian process is realized over a dense grid to better visualize its spatial structure, even though for simulation it is only realized at soil core locations. Here, it is realized over the entire grid, then grayed out over control cells. The spatial pattern of spreading rates has some short-distance variability }
    \label{fig:spreading}
\end{figure*}

Feedstock is spread on treatment cells with a mean application rate of 3.5 kg/m$^2$ of \emph{dry} material. The application rate has variability and spatial structure determined by an anisotropic theoretical variogram \cite{Hoffimann2018} that represents a moderately streaked/striped pattern potentially produced by spreading equipment. This structure is realized using a Gaussian Process. Three realizations of this spatial structure are shown in Figure \ref{fig:spreading}.

Feedstock is assumed to have a constant bulk density of 1000 kg/m$^2$. It has calcium and magnesium concentrations of 7 \% and 5 \% by mass, respectively, with a simulated 3 \% relative standard deviation for both cation concentrations. Feedstock is assumed to be mixed only to shallow depths no greater than 5 cm.

Baseline soil concentrations for calcium and magnesium are also defined by theoretical variograms and realized by Gaussian Processes. The mean soil concentrations are 0.2 \% and 0.1 \% by mass, for calcium and magnesium respectively. Cation concentrations exhibit spatial correlation with themselves and with each other, with a prescribed cross-correlation coefficient of 0.75. The baseline soil bulk density is 1000 kg/m$^3$, on average, with a standard deviation of 100 kg/m$^3$ and spatial structure also defined by a variogram (and realized by a Gaussian Process).

The dissolution and leaching of calcium and magnesium out of feedstock is modeled with exponential functions, $l_i = 1 - \exp(\lambda t)$. Although this is likely not appropriate for more frequent sampling because of seasonally dependent weathering rates, we appeal to the sampling interval of exactly 1 year, which avoids seasonal effects. The decay parameters for calcium and magnesium are 0.4 and 0.8 yr$^{-1}$. This means, for example, that after 1 year of weathering, the loss fraction for calcium is $1 - e^{-0.4} = 0.33$, or 33 \%. For magnesium, it's about 55 \%. These are mostly arbitrary choices for demonstration and do not necessarily constitute a representation of real dissolution rates.

\begin{figure*}
    \centering
    \includegraphics[width=0.9\textwidth]{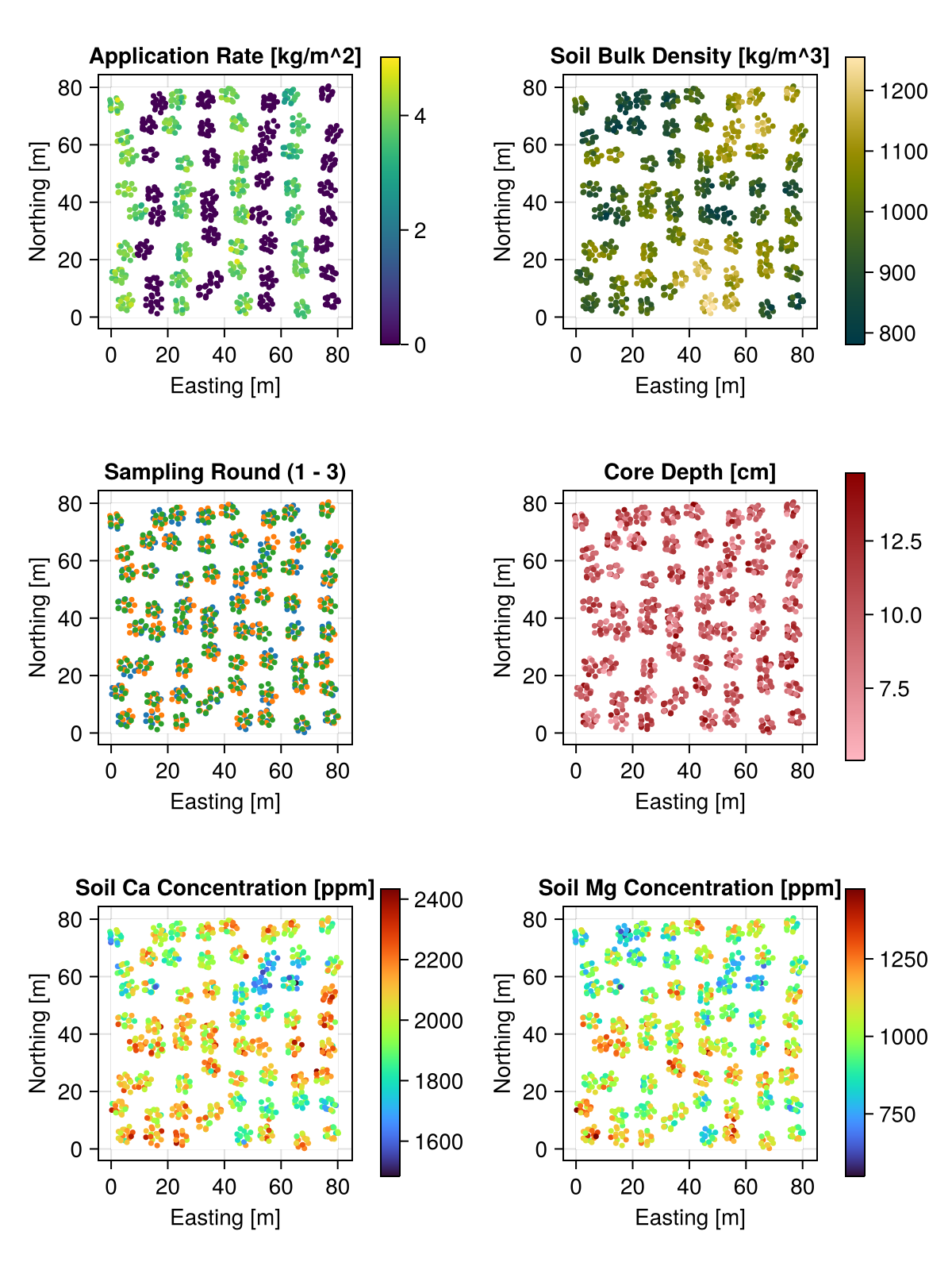}
    \caption{Simulated data for the hypothetical deployment. Each panel shows each of the 960 cores in the sampling plan with their colors defined by various quantities. Each cell of the sampling grid has three circular groups of cores for each sampling round. \textbf{Top left:} Application rates simulated from the anisotropic Gaussian process with control points ignored ($Q=0$). \textbf{Top right:} Soil bulk density. \textbf{Middle left:} The sampling round for each core. There are three rounds of sampling, with the baseline in blue, post-spreading in orange, and post-weathering (1 year) in green. \textbf{Middle right:} Core depth, which varies between 5 and 15 cm. \textbf{Bottom:} Baseline calcium and magnesium concentrations. Each cation is spatially correlated with itself and they are spatially correlated with each other.}
    \label{fig:sim}
\end{figure*}

Figure \ref{fig:sim} shows the values of some key quantities at soil core locations for one realization of the simulation. Note that the application rate, soil concentrations, and soil bulk density exhibit spatial autocorrelation, but sample depth does not. Every simulated quantity is realized at every core in the sampling plan and passed to the mixing model (Section \ref{subsubsec:mixing}) where they are used to compute the exact concentration of the hypothetical soil core pulled from the ground.

\begin{figure}
    \centering
    \includegraphics[width=0.45\textwidth]{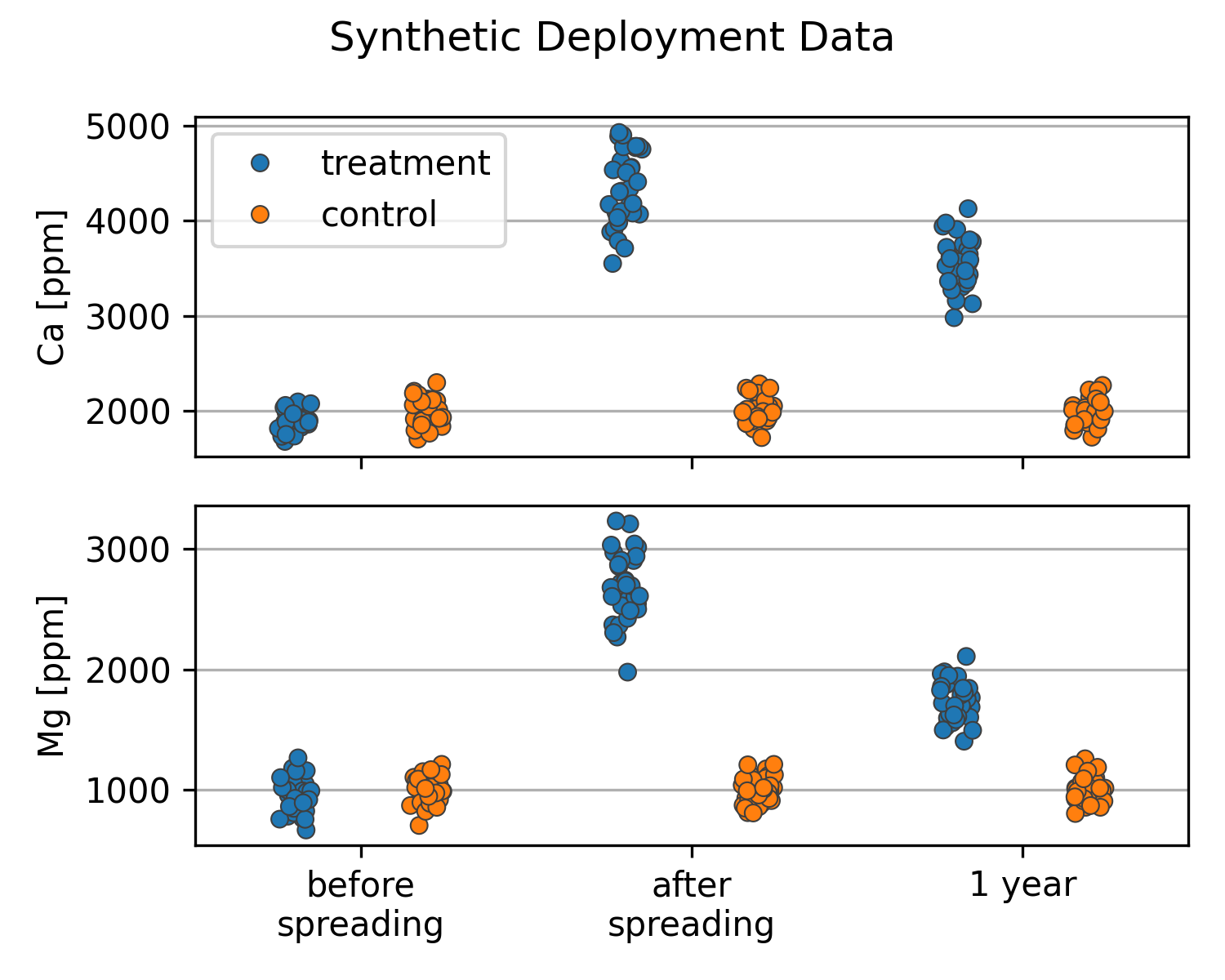}
    \caption{Calcium and magnesium concentrations for the hypothetical deployment in each sampling round. In the ``before spreading" round, treatment and control samples are the same. After spreading, the treatment samples are enriched in both cations and the controls are stable. Then, after weathering, some of the cation enrichment in the treatment samples is lost to weathering.}
    \label{fig:stripplot}
\end{figure}

After compositing each group of cores (averaging the concentrations by mass) analytical noise with 3 \% relative standard deviation is applied. The mass of the final sample is also ``measured" with a 0.5 \% error. Figure \ref{fig:stripplot} shows the final concentrations of each sample, grouped by the sampling round and their treatment/control assignment. As expected, treatment samples are significantly enriched in calcium and magnesium after spreading, and then some of the enrichment is lost to weathering after a year. Control samples have approximately the same concentrations over time.

\subsubsection{Model Definition}
\label{subsubsec:model}

Because samples are taken at approximately the same locations for each sampling round, we model the differences in concentrations over time for the cells of the sample grid. This reduces the impact of spatially variable baseline soil concentrations across the entire plot when sampling different locations for each round. Because the model definition is long, we explain it in pieces, in detail. The model implementation (in PyMC \cite{thomas_wiecki_2024_10821867}) is also \href{https://github.com/LithosCarbon/Monty.jl/blob/5cc0447f69c8621f6e23a9e8e43ff532e6ba7241/cases/demo/demo_module.py}{available online}.

Several deployment and sampling parameters determine the maximum possible amount of initial CDR, which is ultimately set by the total feedstock mass applied. These priors are the total moist feedstock mass applied ($M\wet$), the treatment area ($A$), the sample depth ($d$), the moisture fraction in feedstock ($\phi$), and the feedstock cation concentrations ($\textbf{c}_{f,j})$. We use bold font for symbols representing vectors and bold upper-case for matrices. Note that the subscript $j$ is used for the two cations. The multivariate normal distribution below indicates that the feedstock is, on average, 7 \% calcium and 5 \% magnesium by mass, with 5 \% relative standard deviation for each characterizing the uncertainty. We could, with good reason, also encode a correlation between feedstock cation concentrations by setting the off-diagonal components of the covariance matrix appropriately.
\begin{align*}
    M\wet &\sim \textrm{Normal}(12800, 100) \\
    A &\sim \textrm{Normal}(3200, 32) \\
    d &\sim \textrm{Gamma}(\mu=0.1, \sigma=0.025) \\
    \phi &\sim \textrm{Normal}(0.125, 0.025) \\
    \textbf{c}_{f,j} &\sim \textrm{MvNormal}\left( 
        \begin{bmatrix}
            0.07\\
            0.05
        \end{bmatrix} , 
        \begin{bmatrix}
            0.0035^2 & 0 \\
            0 & 0.0025^2
        \end{bmatrix}
    \right) \\
\end{align*}
From these parameters, we compute the dry feedstock mass ($M_{\textrm{dry}}$) and application rate ($Q$) deterministically.
\begin{align*}
    M\dry &= (1 - \phi) M\wet \\
    Q &= M\dry / A
\end{align*}
The prior for soil density ($\rho_s$) can be based on existing agronomic measurements or an understanding of the soil type. It could also be directly measured for each sample, in which case it would not have a prior and would be incorporated into the model as data. Here we have a prior, which enables subsequent deterministic calculation of the soil mass in the sampled layer ($M\soil$), the expected mixing fraction of feedstock in soil after spreading ($\alpha$), and the \emph{mixed} concentration of each cation in the feedstock-soil mixture after spreading ($\textbf{C}_{\textrm{mixed},i,j}$). Note that we have mixed concentrations for each of $i$ samples and for each of the $j$ cations. $\textbf{C}_{\textrm{mixed},i,j}$ is a two-dimensional array (a matrix) with samples in the rows and cations in the columns. It depends on the mixing fraction, feedstock concentrations, and on the \emph{observed} baseline concentrations $\textbf{C}_{1,i,j}$, where the subscript 1 indicates the first round of samples (taken before spreading). 
\begin{align*}
    \rho_s &\sim \textrm{Normal}(1000, 100) \\
    M\soil &= \rho_s d \\
    \alpha &= Q / (Q + M\soil) \\
    \textbf{C}_{\textrm{mixed},i,j} &= \alpha \textbf{c}_{f,j} + (1 - \alpha) \textbf{C}_{1,i,j}
\end{align*}
The matrix of mixed concentrations represents the expected concentrations after spreading for each sampling location. To compare this with observations, we compute the expected enrichment ($\textbf{E}_{i,j}$) and ``compare" this with the \emph{observed} enrichment between the second and first sampling rounds.
\begin{align*}
    \textbf{E}_{i,j} &= \textbf{C}_{\textrm{mixed},i,j} - \textbf{C}_{1,i,j} \\
    \boldsymbol{\sigma}_{\textbf{E},j} &\sim \textrm{Exponential}(0.001) \\
    \textbf{C}_{2,i,j} - \textbf{C}_{1,i,j} &\sim \textrm{Normal}(\textbf{E}_{i,j}, \boldsymbol{\sigma}_{\textbf{E},j})
\end{align*}
The last line above contains the first likelihood function in the model. It evaluates the probabilities of our observed enrichments, given the expected enrichments for each sample location. The standard deviation vector $\boldsymbol{\sigma}_{\textbf{E},j}$ has two elements, one for each cation.

Now we take care of the controls. In this case, we model the difference in control concentrations between spreading and our observations after one year. To do so, we model the average difference in control samples. We use the concentrations at sampling round 3 and the \emph{average} of sampled concentrations in rounds 1 and 2, which were taken at almost the same point in time. We could alternatively only use observations from round 2, but this would discard relevant information. We use $\boldsymbol{\Omega}$ to denote matrices of control sample concentrations.
\begin{align*}
    \boldsymbol{\sigma}_{\boldsymbol{\omega},j} &\sim \textrm{HalfNormal}(0.001) \\
    \Delta \boldsymbol{\omega}_{j} &\sim \textrm{Normal}(0, 0.001) \\
    \Delta \boldsymbol{\Omega}_{i,j} &= \boldsymbol{\Omega}_{3,i,j} - (\boldsymbol{\Omega}_{1,i,j} + \boldsymbol{\Omega}_{2,i,j})/2 \\
    \Delta \boldsymbol{\Omega}_{i,j} &\sim \textrm{Normal}(\Delta \boldsymbol{\omega}_{j}, \boldsymbol{\sigma}_{\boldsymbol{\omega},j})
\end{align*}
This block of definitions gives us the average change in control sample concentrations over the weathering period ($\Delta \boldsymbol{\omega}_{j}$) for both cations. The prior for these changes is relatively wide compared to average soil concentrations. The last line above contains the second likelihood function in the model. 

Next, we model the change in concentrations due to weathering between sampling rounds 2 and 3. We have already defined parameters representing cation enrichment from spreading and cation changes in the control samples. Now we model losses of calcium and magnesium in the treatment group.

We approach this with a hyperprior for the fraction of cation mass lost from feedstock across both elements. This encodes a loose expectation that calcium and magnesium losses are similar in magnitude. We generally don't expect to see 8 \% calcium loss and 80 \% magnesium loss, for example. This expectation depends on the mineralogy and chemistry of a given feedstock and the parameters chosen here are for demonstration.
\begin{align*}
    \mu_\textrm{loss} &\sim \textrm{Uniform}(0,1) \\
    \sigma_\textrm{loss} &\sim \textrm{Beta}(1, 6) \\
    \boldsymbol{l}_j &\sim \textrm{Normal}(\mu_\textrm{loss}, \sigma_\textrm{loss})
\end{align*}
The average loss parameter $\mu_\textrm{loss}$ represents the mean loss fraction of both calcium and magnesium. It's uniform over the [0,1] interval, which means that we have no particular expectation for cation loss, just that it must be within physical bounds. The model has no prior preference for 0 \% cation loss, 100 \% cation loss, or anything in between. Again, this is just a demonstration, and specific prior expectations for cation losses can be represented.

The parameter $\sigma_\textrm{loss}$ defines how similar we expect the loss fractions to be and the choice above encodes only a loose expectation of similarity. From these hyperpriors, we define two loss fractions $\boldsymbol{l}_j$ for the individual cations.

Now we define two more deterministically produced matrices of concentrations. The loss fractions for each cation ($\boldsymbol{l}_j$) are applied to the modeled enrichment ($\textbf{E}_{i,j}$) and then incorporated into the expected concentrations after weathering ($\textbf{C}_{\textrm{weathered},i,j}$). The post-weathering concentrations account for the average change in concentrations in the control samples ($\Delta \boldsymbol{\omega}_{j})$. 
\begin{align*}
    \textbf{C}_{\textrm{loss},i,j} &= \boldsymbol{l}_j \textbf{E}_{i,j} \\
    \textbf{C}_{\textrm{weathered},i,j} &= \textbf{C}_{2,i,j} - (\textbf{C}_{\textrm{loss},i,j} - \Delta \boldsymbol{\omega}_{j})
\end{align*}
The weathered concentrations computed above are what we observed in the third round of sampling. The third and final likelihood function, after defining a prior for the variability, is below.
\begin{align*}
    \boldsymbol{\sigma}_{\textrm{weathered},j} &\sim \textrm{Exponential}(0.001) \\
    \textbf{C}_{3,i,j} &\sim \textrm{Normal}(\textbf{C}_{\textrm{weathered},i,j}, \boldsymbol{\sigma}_{\textrm{weathered},j})
\end{align*}

We can incorporate deterministic nodes in the model for CDR, in several forms, for convenience.
\begin{align*}
    \textrm{CDR}_{\textrm{potential}} &= Q \boldsymbol{c}_{f,j} \cdot \begin{bmatrix}
        2.196 \\
        3.621
    \end{bmatrix} \\[1ex]
    \textrm{CDR} &= Q (\boldsymbol{l}_j \circ \boldsymbol{c}_{f,j}) \cdot \begin{bmatrix}
        2.196 \\
        3.621
    \end{bmatrix} \\[1ex]
    \textrm{CDR}_{\textrm{completion}} &= \textrm{CDR} / \textrm{CDR}_{\textrm{potential}} \\[1ex]
    \textrm{CDR}_{\textrm{total}} &= A \times \textrm{CDR} / 1000
\end{align*}
The vectors of literal numbers above are the conversion factors for calcium and magnesium, converting their mass losses to \cotwo mass loss for initial CDR (see Equation \ref{eq:cation2co2} and surrounding discussion). The conversion is done by dot product of the two vectors above. The units of CDR for the top two equations are mass per area. CDR completion is a fraction. The total CDR is in metric tons of \cotwo.

There are several modifications to this model that make sense in different scenarios. For example, the likelihood functions can also be T distributions if we'd like the model to be more robust to potential outliers. Cation losses (and the related hyperpriors) can also be defined in a number of ways to encode prior expectation about loss fractions for different scenarios.

\subsubsection{Model Results}
\label{subsubsec:results}

Now we fit the model (Section \ref{subsubsec:model}) using one realization of the hypothetical deployment data (Section \ref{subsubsec:hypodepo}). The realization chosen for this example is the same one shown in Figure \ref{fig:stripplot}. It was the first one produced in a batch of 10,000 realizations of the hypothetical deployment and was not chosen because the results look particularly good or bad by random chance.

We perform prior predictive simulations to characterize and visualize prior distributions and implied distributions. Then we sample from the posterior using the NUTS \cite{hoffman2014no} sampler built into the PyMC package \cite{thomas_wiecki_2024_10821867}. We sample 8,000 draws from 16 chains, discarding the first 4,000 samples from each chain and keeping 64,000 samples total. Sampling completed with zero divergent transitions, no sign of poorly mixing chains, and a maximum $\hat{R}$ statistic \cite{vehtari2021rank} across all parameters of 1.0006. We also check that posterior predictive simulations align adequately with the observations. See \href{https://github.com/LithosCarbon/Monty.jl/tree/75fd398b58451e82599f55f7fa70b744b0cbe236/cases/demo}{the provided code} for more details.

\begin{figure*}
    \centering
    \includegraphics[width=\textwidth]{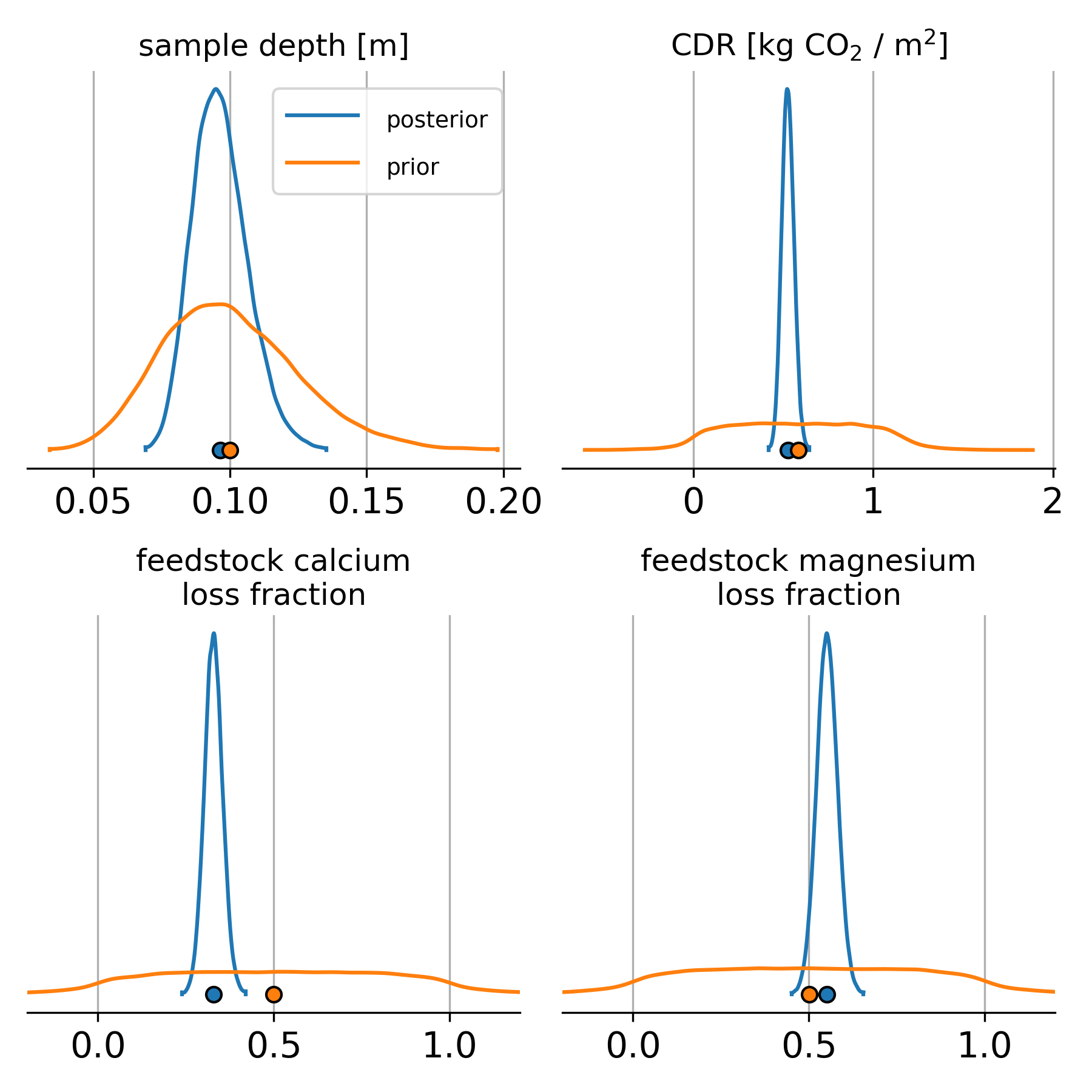}
    \caption{Comparison of prior (orange) and posterior (blue) distributions for a few key model parameters. The dots below each distribution show mean values. The prior for the sample depth is a Gamma distribution with a mean of 10 cm and standard deviation of 2.5 cm based on our understanding of (imaginary) soil core collection. After conditioning the model on the synthetic data, the sample depth is further constrained. The prior for CDR, per unit area, is approximately uniform between zero and the maximum amount possible given applied feedstock mass (in this case about 1 kg/m$^2$). Feedstock cation losses, expressed as fractions of the total mass, are also constrained by prior distributions to be between zero and one with some very limited probability density outside of that interval. Posterior distributions for cation loss fractions narrow considerably, compared to the priors}
    \label{fig:priorposterior}
\end{figure*}

Figure \ref{fig:priorposterior} compares prior and posterior probability distributions for a few key model parameters, using kernel density estimates from the sampled values. The sample depth is updated considerably and the loss fractions for calcium and magnesium contract substantially around mean values of about 33 \% and 55 \%, respectively. This is strong evidence that the model is recovering correct information from the simulated data because the prescribed loss fractions are 33 \% and 55 \%.

\begin{figure*}
    \centering
    \includegraphics[width=\textwidth]{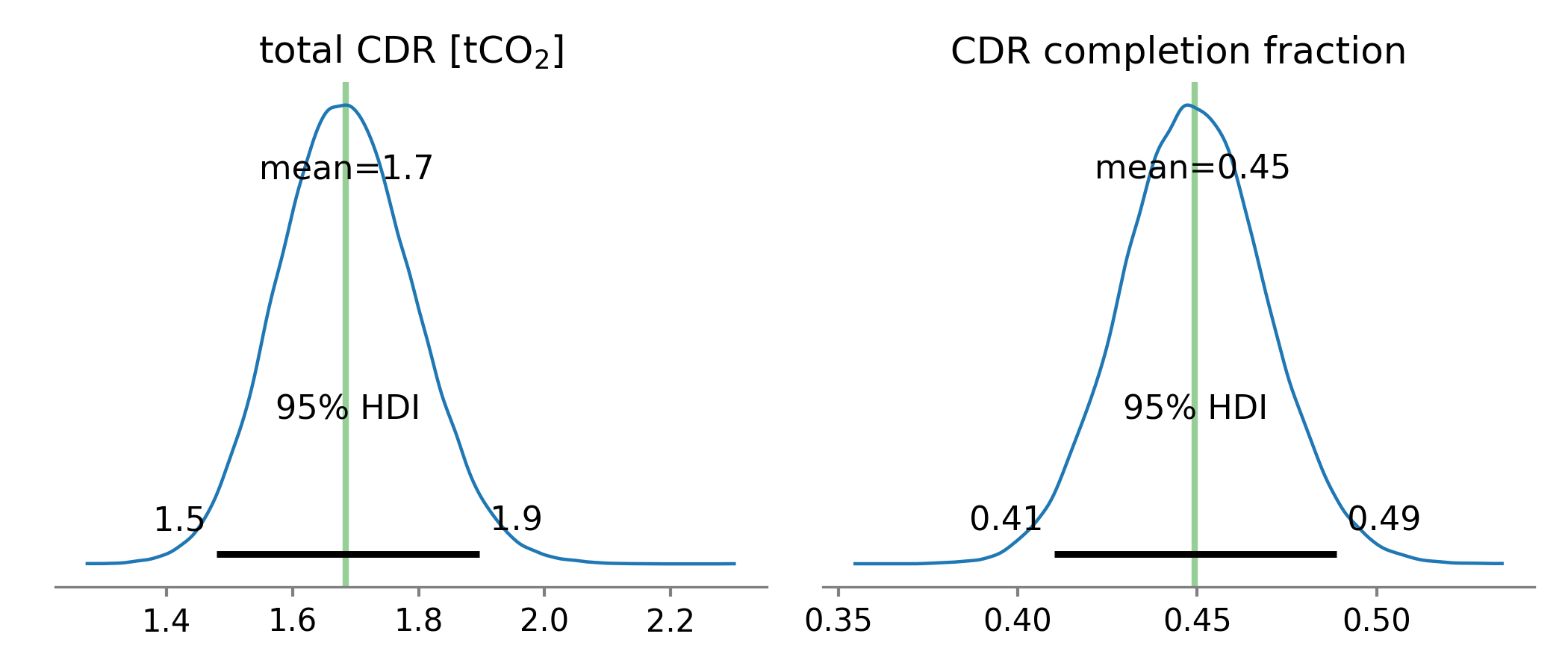}
    \caption{Posterior distributions for initial CDR. Total CDR, in metric tons of CO$_2$, is shown on the left. The CDR completion fraction is shown on the right, which references the maximum amount of initial CDR possible, given the mass of feedstock applied and its cation concentrations. The vertical green line in each panel shows the correct answer, which we know because we prescribed it in the generation of the synthetic data. }
    \label{fig:posteriors}
\end{figure*}

Figure \ref{fig:posteriors} shows posterior distributions for the quantity of greatest interest, initial CDR, along with mean values and 95 \% highest density intervals (HDI). In this case, our inference is that initial CDR is between 1.5 and 1.9 metric tons of \cotwo with 95 \% probability. Equivalently, there is a 95 \% probability that between 41 and 49 \% of the theoretical maximum amount of initial CDR has occurred in the simulated deployment. We know the correct answer for initial CDR because it was prescribed in the \texttt{Monty} simulations. The vertical green lines show the correct values, which are comfortably inside the posterior distributions.

\subsection{Bootstrapping Validation}
\label{subsec:boot}

As mentioned in Section \ref{subsubsec:results}, the simulated deployment described in Section \ref{subsubsec:hypodepo} was realized 10,000 times, drawing randomly from all of the stochastic parameters simultaneously for each realization. Only the first realization was used in the Bayesian model.

\begin{figure*}
    \centering
    \includegraphics[width=\textwidth]{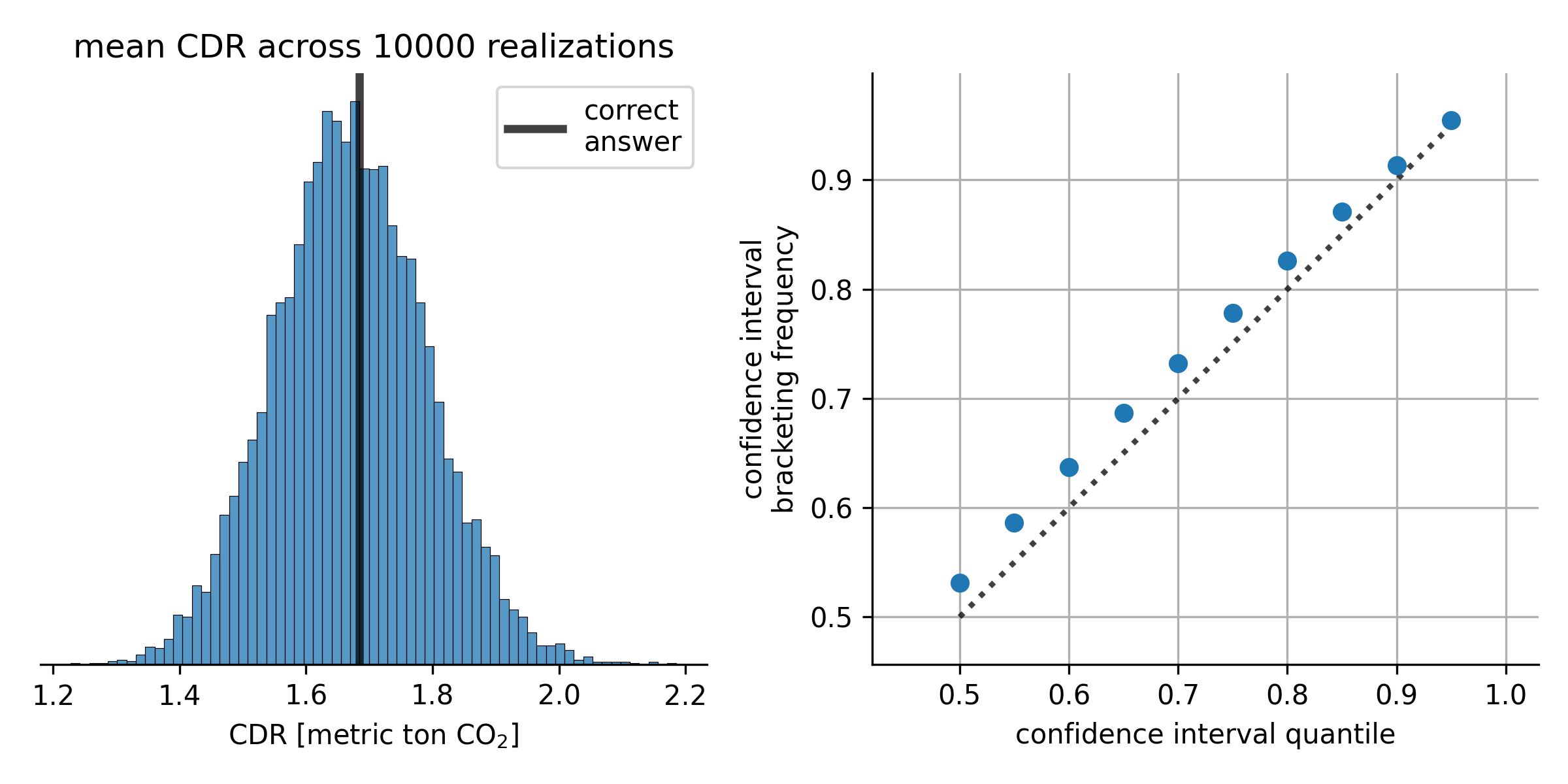}
    \caption{Results from bootstrapped CDR estimation using 10,000 separate realizations of the hypothetical deployment described in Section \ref{subsubsec:hypodepo}. The bootstrapping code for these calculations was written before \texttt{Monty}. Point estimates for each realization are shown in the histogram on the left. Across realizations, the estimates are recovering the correct answer, on average. The panel on the right shows whether bootstrapped confidence intervals are achieving their nominal coverage frequency. The blue dots show how frequently confidence intervals for different quantiles bracket (cover) the correct answer. If so, they should fall along the dotted line. They largely do, indicating that the confidence intervals are correctly representing uncertainty in the estimates.}
    \label{fig:bootstrap}
\end{figure*}

Additionally, we applied our internal analysis pipeline to each of the 10,000 realizations, using non-parametric bootstrapping to estimate initial CDR. This analysis code was written before \texttt{Monty} was created and in Python. Results are shown in Figure \ref{fig:bootstrap}. Bootstrapping recovers the correct answer, on average. Confidence intervals achieve their nominal coverage frequency, bracketing the correct answer at rates close to expectations. This is strong evidence that the code is correct and the analysis pipeline is reliable.

\section{Conclusion \& Summary}
\label{sec:conclusion}

In Section \ref{sec:mass-balance}, we reviewed background information relevant to ERW and explained the goals of mass-balance MRV. Then we detailed the assumptions and equations involved in tracer methods, where immobile elements are used to infer cation losses from feedstock in soil without directly measuring initial concentrations in the mixture. Tracer methods can be powerful but are also fragile in some contexts. First, they rest on assumptions about mass-conservation that are only approximations, which can introduce bias and error. Second, tracer methods break down when the concentration gap between feedstock and baseline soil is too small because the signal is obscured by many sources of variability and noise. It is always better to have higher tracer concentrations in feedstock and tracer methods will not work if feedstock concentrations are \emph{lower} than baseline soil. Third, using multiple tracer elements simultaneously can be advantageous, but different baseline concentration magnitudes must be addressed to fully utilize each element. Finally, tracer elements may not be available. Feedstock may have zero tracer elements with high enough concentrations, given target deployment soils. In general, all of these issues are avoided when tracer concentrations in feedstock are much higher than in baseline soil. Generally, our recommendation is about an order of magnitude.

In Section \ref{sec:mass-balance}, we also discussed tracer-free mass-balance methods, or ``cation stock monitoring" (Section \ref{subsec:cationstocks}). This approach is conceptually and statistically simple, compared to tracer methods. The masses of base cations like calcium and magnesium are monitored over time by measuring their concentrations in soil, along with bulk density and a well-known sampling depth. These elements are always highly enriched in feedstock because they define a feedstock's CDR potential. This approach is also directly analogous to SOC monitoring and applied research in that area can be utilized. Although conceptually simple, cation stock monitoring requires precise operations in the field. In particular, sampling depth must be well-planned, accurate, and consistent across different rounds of sampling.

Section \ref{sec:mass-balance} closes with some discussion of signal size, variability, and noise. Mass-balance methods rely on accurate resolution of concentration changes in soil. For major cations like calcium and magnesium, relevant concentration changes are probably around 100-1000 ppm. For tracer elements, concentration changes could be similar in magnitude but may also be very small for trace elements. These changes, however, are only meaningful in the context of background variability and noise. We discuss this topic in Section \ref{sec:variability}, outlining sources of variability in mass-balance ERW and noise in laboratory analysis.

In Section \ref{sec:planningexperiments}, we review some difficulties and issues in ERW research, which demands rigorous and unbiased quantification of CDR. In general, experiments and field trials need to be fully planned \emph{before} they are carried out, including a clear plan for how collected data will be analyzed. Thorough planning helps ensure that all necessary information is collected at the correct time, that enough samples are collected, and that the analysis plan will address the central research question confidently and accurately. Thorough planning also helps avoid potential biases due to analysis choices, like which groups of data to include/exclude, how to average across groups, which statistical methods to use, etc. Analyzing a dataset in multiple ways, and then presenting only one way that yields a desired result, is not a reliable practice.

One way of testing experimental plans is by simulation of the data-generating process. In Section \ref{sec:simulation} we present a publicly available software package designed to simulate geochemical datasets collected for mass-balance MRV in ERW. The package can be used for several tasks, like understanding how various sources of uncertainty contribute to uncertainty in CDR, validating statistical methods, and selecting appropriate sample sizes. At the core of this package is a simple mixing model, describing how the concentration of an individual soil sample depends on the properties of baseline soil and (potentially) weathered feedstock. In Section \ref{subsec:sensitivity}, we show how this mixing model can be used to understand the sensitivity of a single sample to changes in feedstock and soil properties.

In Section \ref{sec:estimationinference} we review some ways of quantifying uncertainty in initial CDR and address strengths and weaknesses, before discussing some advantages of Bayesian inference in mass-balance MRV. We generate synthetic geochemical data for a hypothetical deployment, present a slightly simplified Bayesian model and infer initial CDR from the dataset. Because we generated the data, we know the correct answer and can check that the model recovers the correct answer, which it does. Finally, in Section \ref{subsec:boot}, we show an example validation of CDR estimates and bootstrapped confidence intervals for the same hypothetical deployment.

\section*{\large Software \& Data Availability}
The software described in this manuscript is open-source and publicly available under the GPL-3.0 license.
\begin{itemize}
    \item The simulation package (\texttt{Monty}) is available at \newline \href{https://github.com/LithosCarbon/Monty.jl}{\small\texttt{https://github.com/LithosCarbon/Monty.jl}}
    
    \item Documentation and examples are available at \newline \href{https://lithoscarbon.github.io/Monty.jl}{\small\texttt{https://lithoscarbon.github.io/Monty.jl}}
    
    \item Simulation results and the results of Bayesian inference on simulated data, as described in Section \ref{subsec:demo_model}, are fully available, in addition to the exact version of the code used to produce these results. The simulation results are archived as both NetCDF files and CSV files, for convenience. Prior predictive and posterior sampling is archived in NetCDF format. These files are all archived at \newline \href{https://doi.org/10.5281/zenodo.11621611}{\small\texttt{https://doi.org/10.5281/zenodo.11621611}}
\end{itemize}

\section*{\large Author Contributions}
\textbf{Mark Baum}: Conceptualization, Data Curation, Formal Analysis, Methodology, Software, Visualization, Writing - Original Draft Preparation.
\textbf{Henry Liu}: Conceptualization, Supervision, Writing - Review \& Editing. \textbf{Lily Schacht}: Conceptualization, Writing - Review \& Editing.
\textbf{Jake Schneider}: Writing - Review \& Editing.
\textbf{Mary Yap}: Supervision.

\printbibliography

\end{document}